%% file: main.tex
\documentclass[12pt]{article} 

\usepackage[utf8]{inputenc}    
\usepackage[T1]{fontenc}       

\usepackage{lmodern}           
\usepackage{booktabs}

\usepackage{colortbl}

\usepackage[utf8]{inputenc}

\usepackage{amsfonts}
\usepackage{amssymb}

\usepackage{array}
\usepackage{longtable}
\usepackage{siunitx}
\usepackage{chemformula}

\usepackage{multirow}

\usepackage{setspace}          
\usepackage[margin=1.00in]{geometry} 
\usepackage{enumitem}
\usepackage{authblk}           
\usepackage{graphicx}          
\usepackage{subcaption}        
\usepackage{amsmath}           
\usepackage{lineno}            
\usepackage{float}             
\usepackage{hyperref}          
\usepackage{listings}          
\usepackage{pdflscape}         
\usepackage[percent]{overpic}  
\usepackage{adjustbox}
\usepackage{algorithm}
\usepackage{algpseudocode}

\usepackage{xcolor}
\definecolor{llmcolor}{RGB}{51, 102, 153}
\definecolor{processcolor}{RGB}{34, 139, 34}
\definecolor{commentcolor}{RGB}{128, 128, 128}

\algrenewcommand\algorithmicrequire{\textbf{Input:}}
\algrenewcommand\algorithmicensure{\textbf{Output:}}
\algrenewcommand\textproc{\textsc}

\usepackage{booktabs}

\definecolor{bgcolor}{rgb}{0.95,0.95,0.95}
\definecolor{commentgray}{rgb}{0.3,0.5,0.3}
\definecolor{keywordblue}{rgb}{0.1,0.1,0.8}
\definecolor{stringred}{rgb}{0.6,0.1,0.1}

\usepackage[numbers,sort&compress]{natbib}

\AtBeginDocument{\setstretch{1.25}}

\begin{document}

\vspace{-2cm}
\title{\vspace{-0.5cm}Automated Machine Learning Pipeline: Large Language Models-Assisted Automated Dataset Generation for Training Machine-Learned Interatomic Potentials.}

\author[1]{\underline{Adam Lahouari}*}
\author[1,2]{Jutta Rogal}
\author[1,3,4,5,6]{Mark E. Tuckerman}

\affil[1]{NYU, Department of Chemistry, New York, NY 10003, USA}
\affil[2]{Initiative for Computational Catalysis, Flatiron Institute, NY 10010, USA}
\affil[3]{NYU, Department of Physics, New York, NY 10003, USA}
\affil[4]{Courant Institute of Mathematical Sciences, NYU, NY 10012, USA}
\affil[5]{NYU-ECNU Center for Computational Chemistry, Shanghai 200062, China}
\affil[6]{Simons Center for Computational Physical Chemistry, NYU, NY 10003, USA}
\affil[*]{Correspondence: al9500@nyu.edu}

\maketitle
\vspace{-1.0cm}


\begin{abstract}
Machine learning interatomic potentials (MLIPs) have become powerful tools to extend molecular simulations beyond the limits of quantum methods, offering near-quantum accuracy at much lower computational cost. Yet, developing reliable MLIPs remains difficult because it requires generating high-quality datasets, preprocessing atomic structures, and carefully training and validating models.  
In this work, we introduce an Automated Machine Learning Pipeline (AMLP) that unifies the entire workflow from dataset creation to model validation. AMLP employs large-language-model agents to assist with electronic-structure code selection, input preparation, and output conversion, while its analysis suite (AMLP-Analysis) based on ASE supports a range of molecular simulations. The pipeline is built on the MACE architecture and validated on acridine polymorphs, where with a straightforward fine-tuning of a foundation model mean absolute errors of ~1.7 meV/atom in energies and ~7.0 meV/\AA\ in forces are achieved. The fitted MLIP reproduces DFT geometries with sub-\AA\ accuracy and demonstrates stability during molecular dynamics simulations in the microcanonical and canonical ensemble.
\end{abstract}

\clearpage

\section{Introduction}
\label{sec:introduction}

Molecular dynamics (MD) simulations are primarily employed to investigate finite-temperature effects, dynamical properties, and time-correlation functions in materials.\cite{Md-headtransfer,MD-temperature-elastic} While classical force fields enable access to long time and length scales, their inherent approximations can limit the accuracy of these predictions.\cite{limitation-MD} These classical force fields are often parameterized or fitted to specific types of systems or interactions, leading to inadequacies when attempting to generalize their use to broader, more complex scenarios. Particularly problematic is their inability to reliably distinguish subtle energy differences between closely related polymorphs, which typically differ by only a few kJ/mol.\cite{md-poly1,md-poly2} Such precision is critical in determining material stability, phase transitions, and reactive pathways, highlighting the need for methods that can accurately describe these delicate energy landscapes.

Quantum mechanical (QM) methods, such as Density Functional Theory (DFT) and post-Hartree-Fock approaches, offer high accuracy in capturing electronic structure details.\cite{qc-expense} However, their computational expense rapidly escalates with system size, making them impractical for routine simulations of large-scale molecular systems simulations. This computational constraint limits their applicability to small or medium-sized systems.

In response to these limitations, machine learning interatomic potentials (MLIPs)  have emerged as a robust alternative, bridging the accuracy of quantum chemical methods and the computational efficiency of classical force field potentials.\cite{QC-ML1,QC-ML2} MLIPs learn from quantum chemical data, providing high-fidelity representations of atomic interactions while remaining computationally efficient enough to handle large systems and dynamical simulations over longer timescales. They offer substantial improvements in both predictive accuracy and scalability, facilitating precise simulations of complex systems under a wide variety of conditions. However, having these benefits depends on the quality and comprehensiveness of the training dataset. 

The key factor in developing an accurate model capable of effectively predicting user requirements lies in constructing a precise dataset. Initially, careful consideration of the underlying chemistry of the system is essential, particularly identifying the dominant interactions influencing system stability, such as long-range interactions for example. Subsequently, selecting the appropriate QM method is crucial. While DFT remains the most common training data source for MLIPs due to its efficiency and moderate computational cost, higher-level QM methods, including CCSD~\cite{ccsdpaper-reviewer1} and multi-reference approaches,\cite{multireference-reviewer1} can be used as the target level of theory.

Furthermore, it is crucial to ensure that the dataset contains structural configurations extending beyond the equilibrium minima. Typically, this sampling procedure begins with geometry optimization (or cell optimization in the case of crystals systems), followed by ab initio molecular dynamics (AIMD) simulations conducted at various temperatures. In principle, such a protocol enables the dataset to sample a broad region of the potential energy surface (PES), capturing not only near-equilibrium configurations but also thermally accessible distortions, anharmonic vibrations, and local structural rearrangements. In practice, however, the extent of PES coverage is constrained by the timescales and temperature ranges explored during AIMD. For example, simulations at moderate temperatures primarily sample configurations near the equilibrium minima, whereas high-temperature trajectories can probe a wider basin of the PES, although at the cost of introducing highly distorted structures that may not be representative of typical conditions. The inclusion of these diverse structural samples significantly enhances the MLIP’s ability to generalize and adapt effectively across different simulation regimes, especially in MD scenarios. Conversely, a limited or inadequately constructed dataset covers only a narrow region of the PES, which can lead the model to learn spurious correlations and produce unrealistic or unreliable predictions outside the training domain. Standardized datasets now support evaluation across distinct domains—for example, Matbench for inorganic materials, Matbench Discovery for large-scale crystal-stability assessment, and OMC25 for molecular crystals.\cite{mathbench,MatbenchDiscovery,omc25} However, for system-specific training, these datasets must be expanded. To address this, our Automated Machine Learning Pipeline (AMLP) is designed to function as a ``data engine'' within a human-driven active-learning (AL) framework for MLIPs. This design follows established users total control over each step of the AL cycle. The result is a portable, adjustable workflow—transferable across new structures and DFT codes, from data selection through final validation.
To make this workflow accessible to non-experts, we integrated an LLM-powered multi-agent system that provides guidance on system-specific problems. This approach, which is under intense development in the community, has shown promise in solving direct quantum-chemistry tasks and facilitating ad hoc problem solving.\cite{ZOU2025102263}
Once a dataset is assembled, model selection and hyperparameter optimization become central to reliable training and fair comparison. For the training stage, MACE is a strong choice: it offers pretrained foundation models spanning broad chemical space with a design that captures fine-grained atomic environments, delivering strong baseline accuracy while markedly reducing data and compute requirements during task-specific fine-tuning \cite{MACE,fine-tuning-juttafriend}.

As training a potential from scratch can be highly resource-intensive and time-consuming, foundational models have become increasingly prevalent. These models are pre-trained on extensive datasets derived from thousands of DFT calculations. By leveraging these pre-trained checkpoints, users can efficiently fine-tune models to their own datasets, benefiting from improved accuracy due to the foundational model's existing chemical insights. Additionally, fine-tuning typically achieves faster convergence with fewer training epochs, requires less training data, and ensures robust performance due to prior exposure to diverse chemical environments.
Despite these substantial advancements, the training, validation, and deployment of MLIPs continue to pose significant technical and practical challenges. The process typically demands extensive effort, as each aspect—from dataset preparation to model training and validation—requires attention and specific data formats compatible with various software packages.

To address these challenges, we introduce a comprehensive pipeline called Automated Machine Learning Pipeline (AMLP). AMLP simplifies the creation of MLIPs from straightforward input files, such as Crystallographic Information Files (\texttt{.cif}) or Cartesian coordinate files (\texttt{.xyz}). AMLP comprises five distinct steps, each designed to streamline the overall process. Moreover, AMLP leverages Large Language Models (LLMs) to assist users in identifying suitable QM methodologies for their systems. In practice, the user provides a concise prompt describing the target system, and the LLM queries its knowledge of the published literature to suggest methods, basis sets, or dispersion corrections that have been successfully applied to similar cases. This capability not only offers an informed starting point for parameter selection but also helps the user rapidly survey prior research, thereby reducing the need for extensive manual literature review and enabling faster setup of reliable QM calculations. Subsequently, AMLP facilitates the automated creation of input files necessary for geometry optimization and AIMD simulations compatible with popular computational software such as Gaussian\cite{g16}, VASP\cite{vasp1,vasp2,vasp3,vasp4}, CP2K\cite{cp2K}, Orca\cite{orca}, and Psi4\cite{psi4}.

Following simulations, AMLP systematically processes outputs into structured \texttt{.json} files, enabling convenient data storage, retrieval, and visualization of pertinent structural information. Finally, AMLP converts these processed datasets into HDF5 format, thereby facilitating seamless integration with ML training frameworks like MACE, further streamlining and simplifying the entire model-building workflow.

The resulting potentials can subsequently be utilized through an analysis module, AMLP-Analysis, integrated with the Python-based Atomic Simulation Environment (ASE), facilitating flexible evaluation and simulation tasks.\cite{ase-paper}
AMLP supports several dataset-construction strategies to achieve comprehensive coverage of the PES and minimize unsampled regions. Users may (i) perform cell optimizations followed by AIMD to generate training configurations and train an in-house MLIP; (ii) leverage foundation models to explore configuration space via the AMLP-Analysis module, then compute single-point reference data for the sampled structures using the data-generation module; or (iii) adopt a hybrid, active-learning workflow that begins with low-cost methods and iteratively refines the model with selectively added high-level calculations. In all cases, AMLP provides a unified, user-friendly interface that enables practitioners to adopt any of these strategies with minimal effort.

\begin{figure}[H]
    \centering
    \includegraphics[width=\linewidth]{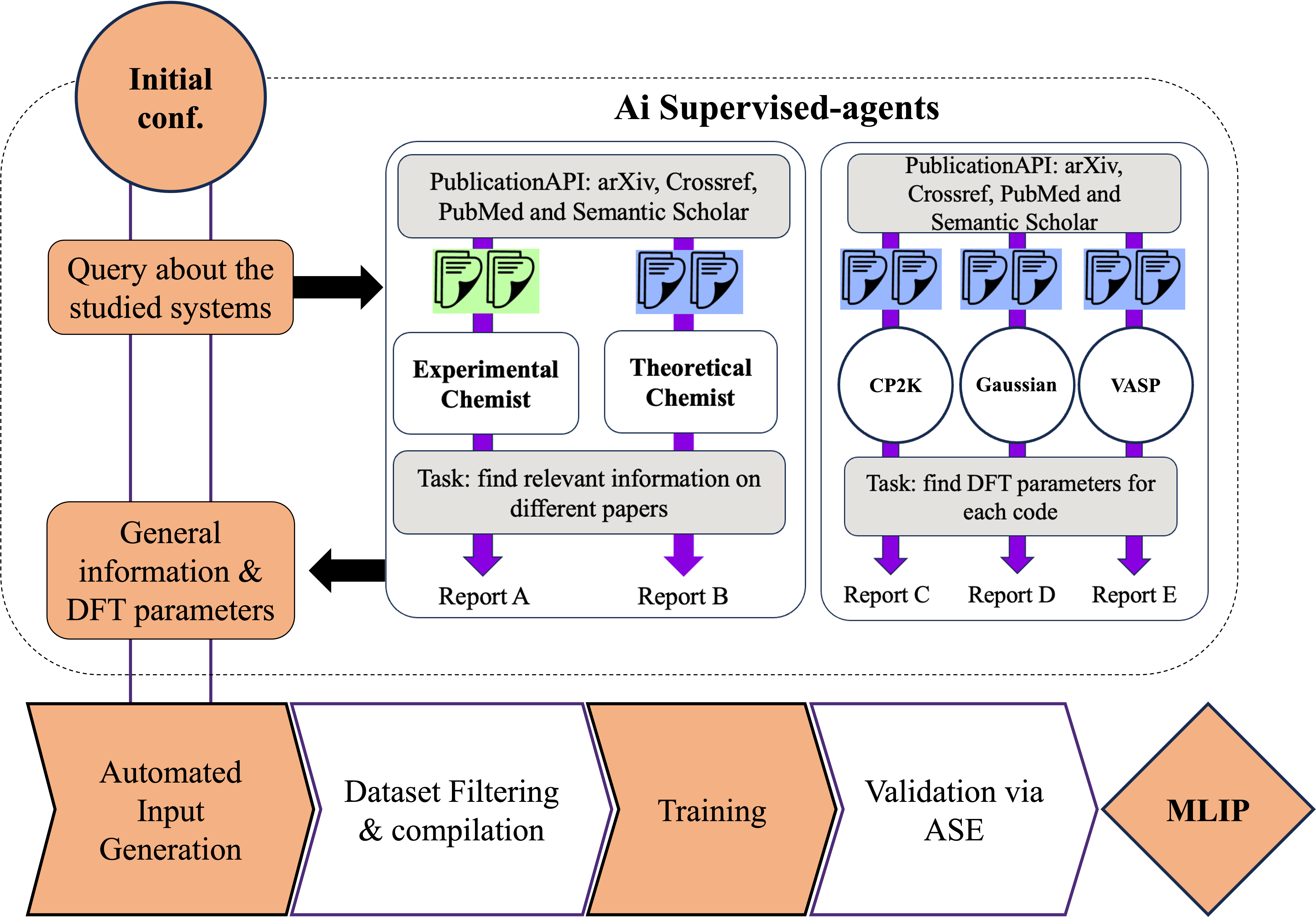}
    \caption{Roadmap of the Automated Machine Learning Potential (AMLP) framework. 
    The workflow begins with structural inputs (\texttt{.xyz}, \texttt{.cif}), which are processed by AI supervised agents to extract relevant literature data and recommend DFT parameters for different codes (e.g., CP2K, Gaussian, VASP). 
    Automated input generation produces ready-to-use files for geometry optimization, cell optimization, single point calculation and AIMD. 
    Simulation outputs are curated into \texttt{.json} datasets containing energies, forces, and structural information, which are then used to train MLIPs. 
    Validation via ASE includes geometry and cell optimization, MD simulations, RDF calculations.
    This roadmap illustrates the fully automated pipeline from raw input structures to validated MLIPs.}
    \label{fig:roadmap}
\end{figure}

\section{Code architecture}

\subsection{General flow chart of AMLP}

The complete AMLP workflow integrates a multi-agent LLM system with automated computational processes to create a pipeline from structure input to machine learning potential deployment. Algorithm~\ref{alg:amlp_main} presents the overall architecture, which consists of two main components: (A) an automated data preparation pipeline that uses the multi-agent LLM recommendations to generate training datasets, and (B) post-training simulation tools that enable immediate application of the trained models. The pipeline is designed to minimize manual intervention.

The data preparation pipeline (Part A) has six sequential steps: structure parsing, LLM-guided method selection, quantum mechanical calculation setup, batch processing of results, automated AIMD generation, and final dataset formatting for MACE training. 

\begin{algorithm}[htb]
\caption{Automated Machine Learning Pipeline (AMLP)}
\label{alg:amlp_main}
\begin{algorithmic}[1]
\Require structure\_file (\texttt{.xyz}/\texttt{.cif}), user\_description, target\_software
\Ensure mace\_dataset, trained\_model\_tools

\State \textcolor{commentcolor}{// PART A: DATA PREPARATION PIPELINE}
\Function{amlp\_data\_pipeline}{}
    \State \textcolor{processcolor}{// Step 0: Structure Processing}
    \State $structure \gets \textproc{parse\_structure\_file}(structure\_file)$
    
    \State \textcolor{llmcolor}{// Step 1: Multi-Agent LLM Method Selection}
    \State $qm\_params \gets \textproc{llm\_method\_recommendation}($
    \State \qquad $user\_description, structure)$
    
    \State \textcolor{processcolor}{// Step 2: Generate Initial Calculations}
    \State $calc\_inputs \gets \textproc{generate\_calculations}(structure, qm\_params)$
    \State \quad $\triangleright$ create single-point, geometry relaxation, cell optimization
    
    \State \textcolor{processcolor}{// Step 3: Process QM Outputs (Batch)}
    \State $qm\_results \gets \textproc{process\_all\_qm\_outputs}(calc\_inputs)$
    
    \State \textcolor{processcolor}{// Step 4: Auto-Generate AIMD Inputs}
    \State $aimd\_inputs \gets \textproc{create\_aimd\_from\_qm}(qm\_results, qm\_params)$
    
    \State \textcolor{processcolor}{// Step 5: Process AIMD \& Convert to MACE Format}
    \State $aimd\_outputs \gets \textproc{process\_aimd\_outputs}(aimd\_inputs)$
    \State $mace\_dataset \gets \textproc{convert\_to\_mace\_format}($
    \State \qquad $aimd\_outputs, qm\_results)$
    
    \State \Return $mace\_dataset$
\EndFunction

\State \textcolor{commentcolor}{// PART B: POST-TRAINING SIMULATION TOOLS}
\Function{simulation\_tools}{trained\_mace\_model}
    \State \Return simulation toolkit for MD, properties, optimization
\EndFunction

\end{algorithmic}
\end{algorithm}

These steps allow users to easily utilize individual tools without needing to repeat the entire procedure. The initial step involves generating detailed feedback reports through AI-driven agents using a single user prompt as depicted in Algorithm~\ref{alg:llm_agents}, providing guidance on suitable QM methods to use and on the parameter settings for the investigated system. 
Subsequently, the second step facilitates automated input file generation, supporting various computational software options (CP2K, VASP, or Gaussian) for geometry or cell optimization. The third step dynamically processes simulation outputs from the chosen software, systematically extracting energies, forces, and coordinates, and formatting these into structured \texttt{.json} files. These \texttt{.json} files are instrumental in identifying global minima configurations, which subsequently enable the automatic creation of input files for AIMD simulations—constituting another key step in the workflow. Finally, by reprocessing both geometry optimization and AIMD simulation outputs, the pipeline generates HDF5 datasets in the format required by the current version of MACE. These files contain the training and validation data in a structure directly readable by the MACE training routines, completing the dataset preparation step. The use of HDF5 allows hierarchical storage of atomic configurations, energies, and forces in a single portable file. This makes the datasets not only compatible with MACE but also adapted for use in other codes that accept HDF5 inputs.

A tutorial that guides users through the general workflow is available online.\cite{colablink-amlp} Even non-expert users can follow the steps to process single \texttt{.cif} files and construct a machine learning potential within a single session.

\subsection{Part A: Data preparation pipeline}
\label{sec:datapreparation}
\begin{enumerate}
    \item \textbf{\textit{AI-agent feedback}}
    
        The training procedure begins with AI-driven agents designed to provide users with tailored insights into their chemical systems, in line with their specific research objectives. These agents also offer an initial suggestion regarding the appropriate DFT code and parameter settings that may be suitable for their study. While LLMs are already being explored in the field of chemistry to generate insights from user queries, this area is still rapidly evolving due to its precision and ease of integration in different codes and workflows.\cite{comparison_model_studens}
        
        Numerous efforts are underway to integrate LLMs into domain-specific computational frameworks.\cite{researchassi,chemgraph} In our approach, we follow a similar paradigm by using an LLM-based model to suggest initial DFT setups. Additionally, we introduce an application of general-purpose AI tools, where queries are progressively refined and delegated to specialized agents tailored.
        
        To use the AMLP framework, an account with an API platform of one of the current LLM providers is needed. Details on the setup with the latest LLM models are regularly maintained and updated in the AMLP repository.
        We recommend using a model that provides a favorable balance between performance and cost-efficiency. The model selection can then be configured by editing a simple \texttt{config.yaml} file. Furthermore, each agent described below can be fine-tuned independently by specifying a distinct model and custom keywords suited to its specific role within the multi-agent framework.
        
        The multi-agent system operates through a structured workflow where specialized agents collaborate to provide comprehensive quantum mechanical method recommendations. Initially, the AI agents engage with users interactively to collect detailed information about their objectives and requirements. The system then has a five-stage process: (1) query refinement to optimize the research question, (2) parallel literature analysis by experimental and theoretical chemistry specialists, (3) scientific integration of findings, (4) consultation with DFT software experts, and (5) synthesis of final recommendations. Algorithm~\ref{alg:llm_agents} details this multi-agent coordination process, showing how each specialized agent contributes to the overall method selection workflow.
        
        \begin{algorithm}[htb]
        \caption{Multi-Agent LLM System for QM Method Selection}
        \label{alg:llm_agents}
        \begin{algorithmic}[1]
        \Require user\_prompt, atomic\_structure
        \Ensure final\_recommendation, literature\_references
        
        \Function{llm\_method\_recommendation}{user\_prompt, atomic\_structure}
            \State $system\_desc \gets \textproc{analyze\_structure}(atomic\_structure) \oplus user\_prompt$
            
            \State \textcolor{llmcolor}{// Initialize Specialized LLM Agents}
            \State $agents \gets \{\textproc{ExpChem}, \textproc{TheoChem}, \textproc{Gaussian},$
            \State \qquad $\textproc{VASP}, \textproc{CP2K}\}$
            \State $supervisors \gets \{\textproc{Integration}, \textproc{DFT\_Rec}\}$
            
            \State \textcolor{llmcolor}{// Step 1: Query Refinement Chain}
            \State $refined\_query \gets \textproc{refine\_query}(user\_prompt, system\_desc)$
            
            \State \textcolor{llmcolor}{// Step 2: Multi-Agent Literature Analysis}
            \State $exp\_analysis \gets agents.\textproc{ExpChem}.\textproc{summarize\_refs}(refined\_query)$
            \State $theo\_analysis \gets agents.\textproc{TheoChem}.\textproc{summarize\_refs}(refined\_query)$
            
            \State \textcolor{llmcolor}{// Step 3: Scientific Integration}
            \State $integrated\_report \gets supervisors.\textproc{Integration}.\textproc{integrate}($
            \State \qquad $exp\_analysis, theo\_analysis)$
            
            \State \textcolor{llmcolor}{// Step 4: DFT Software Expert Consultation}
            \For{$expert \in \{\textproc{Gaussian}, \textproc{VASP}, \textproc{CP2K}\}$}
                \State $expert\_analysis[expert] \gets agents.expert.\textproc{analyze\_refs}(refined\_query)$
            \EndFor
            
            \State \textcolor{llmcolor}{// Step 5: Final Synthesis}
            \State $dft\_content \gets \textproc{concatenate}(expert\_analysis.values())$
            \State $final\_rec \gets supervisors.\textproc{DFT\_Rec}.\textproc{integrate}(dft\_content)$
            \State $all\_refs \gets \textproc{combine\_refs}(all\_agent\_outputs)$
            
            \State \Return $final\_rec, all\_refs$
        \EndFunction
        
        \end{algorithmic}
        \end{algorithm}
        
        The literature review is implemented through the \emph{PublicationAPI} class, which retrieves and analyzes publications from multiple databases, including arXiv\cite{arxiv}, Crossref\cite{crossref}, PubMed\cite{pubmed}, ChemRxiv \cite{chemrxiv} and Semantic Scholar\cite{semantic_scholar}. These sources can be modified in the same configuration file where the agents’ models are defined. Parameter extraction relies primarily on LLM-based methods, with a fallback to regular expressions (Regex) if LLM parsing fails. Regex provides pattern-matching tools that locate, extract, and validate text based on predefined syntax. In this context, Regex serves as a backup for extracting computational details—such as DFT parameters—from scientific literature when AI-based extraction is unsuccessful. 
        
        The framework defines specialized chemistry agents. Two specialized agents are defined. The \emph{ExperimentalChemistAgent} emphasizes keywords related to laboratory practice (experiment, synthesis, characterization, measurement, spectroscopy, microscopy, analysis), while the \emph{TheoreticalChemistAgent} focuses on computational approaches (theory, simulation, computation, calculation, DFT, quantum, molecular dynamics, ab initio, functional). Each agent focuses on its respective domain—computational or experimental chemistry—using a relevance-ranking system. Publications are scored based on keyword matches in titles and abstracts (with weighted scoring), adjusted by a recency bonus for recent works. Review articles are prioritized to ensure broad coverage of the field, and queries are tailored to system-specific contexts (e.g., crystals, metals, organics, polymers). Users may also customize the keyword lists to fine-tune the search criteria for each agent. Finally, each agent compiles its findings into a report tailored to the user’s needs.
        
        Upon approval from the supervisory agent, these refined reports are dispatched to specialized expert agents proficient in the widely utilized computational chemistry codes: VASP, Gaussian, and CP2K. Each of these expert agents provides detailed feedback and recommendations on the applicability and suitability of their respective codes for the user's specific research needs. 
        
        Users first choose their preferred computational code based on expert feedback and receive guidance on appropriate parameter settings thanks to the reports. However, because LLM-based agents may fabricate data when source information is unavailable, a single incorrect output can propagate through the pipeline—creating chained dependencies in which one agent’s error misguides all subsequent agents. To guard against this cascading failure, we have strengthened each individual agent and our supervisory layer by implementing validation checkpoints and refining their prompts, thereby minimizing the risk of generating false information.
    \item \textbf{\textit{Input Generation}}

        Based on the AI recommendations and user-defined parameters, input files are generated from \texttt{.cif} or \texttt{.xyz} formats. Two options are available:
    
        \begin{enumerate}
            \item \textbf{Batch-mode}: Automatically generates inputs for all structures using default templates.
            \item \textbf{Guided-mode}: Allows users to specify DFT parameters interactively.
        \end{enumerate}
        
        For example, in the VASP workflow, AMLP identifies all input structures (e.g. \texttt{.cif} or xyz) and prompts the user to define simulation cell size, functionals, cutoff energy, and other DFT parameters for all of them. The INCAR and KPOINTS files are automatically generated. It allows a step-by-step generation of the inputs for different structures at one time. Though template-driven, users are encouraged to verify all generated parameters. Minor discrepancies may exist depending on software versions, and basic familiarity with each package is recommended.

    \item \textbf{\textit{Output processing}}
        
        Upon obtaining optimized molecular structures, users may proceed with the provided post-processing script, which is designed to extract recursively and organize essential computational outputs into a structured \texttt{.json} format. This includes atomic coordinates, total energies, atomic forces, elements, and simulation cell parameters (including cell lengths, angles, and volumes). Furthermore, the resulting \texttt{.json} file enables users to validate easily whether the optimization has properly converged and to identify any anomalies or errors that may have occurred during the simulation. 
        
    \item \textbf{\textit{AIMD processing}}

        The AIMD setup module begins by preprocessing the simulation output files. It systematically analyzes all \texttt{.json} files within a specified directory, detecting the atomic species in each file individually. For every system, it extracts the final geometry from the previous calculation and generates corresponding VASP or CP2K input files, incorporating user-defined parameters.
    
        As in previous steps, AIMD simulation inputs are tailored to user specifications. A structured questionnaire interface allows users to explicitly define essential simulation parameters such as the target temperatures, integration timestep, and other relevant computational settings. For users less familiar with AIMD setups, the interface offers guidance to ensure that simulation parameters—such as those approaching melting temperatures—are selected appropriately for the chemical system under study.
        
        Upon completion of the AIMD simulations and subsequent evaluation of energies and forces, the output can be processed as with the last step, to produce temperature-specific \texttt{.json} datasets.
        These files serve as structured inputs for MLIP training or further electronic structure investigations.

    \item \textbf{\textit{ML potential dataset creation}}

        The dataset containing all relevant information in a single \texttt{.json} file can be automatically converted into an HDF5 format, which is then partitioned into training and validation subsets—by default using an 85\% to 15\% split, although this ratio can be adjusted by the user. During this process, users are also prompted to define the batch size and specify a maximum threshold for atomic forces. While including moderately ``high-force'' configurations is beneficial for training robust MLIPs (as discussed in Section \ref{sec:introduction}), excessively high-force configurations—often artifacts of insufficient equilibration, numerical instabilities, or unphysical/inaccessible geometries—can corrupt the learning of the PES and lead to poor generalization. The force threshold allows users to identify and exclude such extreme outliers while retaining the diverse, physically relevant configurations that improve model performance.'' Once the dataset is prepared, users can directly launch the training of a MLIP using the MACE architecture. A sample \texttt{config.yaml} file—containing all necessary parameters for training—is available in the project’s GitHub repository and can be adapted by users to suit their specific needs.\cite{amlp_github} This ensures a straightforward transition from dataset generation to model training while preserving flexibility for methodological adjustments.

\end{enumerate}

The user then initiates MACE training using the AMLP-generated HDF5 datasets and configuration templates. Upon successful training, users obtain a customized MLIP, which can subsequently be directly loaded and utilized with the next module.

\subsection{Part B: Post-training simulation tools}
\label{sec:amlpa}
This subsection enables users to load and apply their MLIP for a variety of computational analyses using the AMLP-Analyse (AMLPA) module. The process is streamlined through a straightforward \texttt{config.yaml} configuration file, allowing users to conveniently execute multiple computational tasks simultaneously. This validation module uses ASE, a widely-used Python-based toolkit known for its simplicity, flexibility, and extensive range of features suitable for atomic-scale simulations.\cite{ase-paper}
        
Users have several options for simulation types, including single-point energy calculations, geometry optimizations and cell optimizations. Geometry optimization can be activated by setting the corresponding flag to in the \texttt{config.yaml} file.
Detailed information about these optimizers can be readily accessed through the \href{https://wiki.fysik.dtu.dk/ase/install.html}{ASE documentation}.\cite{ase_website}
Upon successful completion of geometry optimization, the resulting data—including lattice parameters, total energy, simulation cell dimensions and angles, stress tensors, atomic coordinates, and individual atomic forces—are comprehensively reported and systematically saved into an \texttt{.xyz} file for further computational tasks.
        
Subsequent MD simulations can be performed. For simulations in the canonical (NVT) ensemble, users may employ either a Langevin\cite{SchneiderStoll1978}
or a Nos{\'e}-Hoover chain thermostat\cite{Nose1984,Hoover1985,Martyna1992}.
The Langevin integrator is particularly robust in dissipating excess energy and equilibrating systems with rugged potential energy surfaces, while the Nos{\'e}-Hoover chain integrator provides smoother canonical sampling without stochastic noise.  
        
Alternatively, simulations may be conducted in the microcanonical (NVE) ensemble, in which the total energy is conserved. Here, the velocity-Verlet integrator is used to propagate the dynamics. Users can therefore choose between deterministic (Nosé-Hoover chain, velocity-Verlet) and stochastic (Langevin) schemes depending on the desired ensemble and level of control over thermal fluctuations. Energy conservation metrics are computed automatically for NVE runs to verify the stability of the integration. The energy conservation measure is calculated via
\begin{equation}
    \Delta E = \frac{1}{N_{step}} \sum_{k=1}^{N_{step}} \frac{|E_k - E(0)|}{|E(0)|}
\end{equation}
where $E(0)$ is the reference energy at the beginning of the analysis period (after equilibration), $E_k$ is the total energy at step $k$, and $N_{step}$ is the number of simulation steps up to the current point. This formulation provides a cumulative measure of energy conservation quality that accounts for all deviations from the initial energy, normalized by the reference energy.\cite{marksbook}
Additionally, users can replicate the computational cell in multiple spatial dimensions to model larger systems and achieve more comprehensive statistical sampling.
        
From the final MD trajectories, the radial distribution function (RDF) $g(r)$ is computed using the \href{https://mdtraj.org/1.9.4/index.html}{mdtraj} library.\cite{mdtraj} The RDF is defined as
\begin{equation}
    g(r) = \frac{1}{4\pi r^{2}\rho N} \left\langle \sum_{i=1}^{N}\sum_{j\neq i}^{N} \delta\!\left(r - \lvert \mathbf{r}_{i} - \mathbf{r}_{j} \rvert \right) \right\rangle ,
\end{equation}
where $\rho$ is the average number density, $N$ is the number of atoms, and $\langle \cdot \rangle$ denotes an ensemble average. 
Users can specify whether the RDF is calculated globally or selectively for particular atomic pairs, such as nitrogen–nitrogen pairs, and RDFs calculated at different temperatures are collectively presented on a single plot to facilitate direct comparisons against the initial configuration.

This comprehensive computational framework gives users with the tools necessary to develop, rigorously validate, and utilize their MLIP effectively, as summarized in Figure \ref{fig:roadmap}. In the next section, we propose a case study where we exemplify the use of AMLP in order to create an MLIP for a specific crystal structure.

\section {Case study}

\subsection{MLIPs for Acridine}

We apply the new AMLP pipeline to acridine, a molecular crystal known to exhibit nine distinct crystal structures: eight polymorphs and one hydrate.\cite{review_schur} The primary challenge arises from their small lattice energy differences, typically within just a few kJ/mol.\cite{distribution_lattice_energy} Consequently, we seek to develop an MLIP capable of accurately distinguishing among the subtle structural variations of acridine polymorphs, thereby significantly improving predictive reliability compared to empirical force fields.
All pure acridine polymorphs possess monoclinic symmetry, with experimental structures available from the Cambridge Structural Database (CSD) measured at room temperature. Due to variations in polymorph notation across different studies, we consistently adopt the naming convention provided by the CSD, as summarized in Table~\ref{tab:acr_polymorphs}.

\begin{table}[htbp]
  \centering
  \normalsize 
  \setlength{\tabcolsep}{4pt}
  \caption{Crystallographic data for acridine polymorphs}
  \label{tab:acr_polymorphs}
  \begin{tabular}{llrrrrr}
    \toprule
    Polymorph & Space group & \multicolumn{1}{c}{$a$ (\AA)} & \multicolumn{1}{c}{$b$ (\AA)} & \multicolumn{1}{c}{$c$ (\AA)} & \multicolumn{1}{c}{$\beta$ (°)} & \multicolumn{1}{c}{$V$ (\AA$^3$)} \\
    \midrule
    ACRDIN04 & P2$_1$/n & 11.25(1) &  5.95(1) & 13.60(1) & 99.5(1) &  898 \\
    ACRDIN05 & Cc       &  6.17(2) & 23.50(8) & 12.87(4) & 96.5(1) & 1855 \\
    ACRDIN06 & P2$_1$/n &  6.06(1) & 22.81(4) & 13.20(2) & 95.9(1) & 1815 \\
    ACRDIN09 & P2$_1$/c &  6.05(7) & 18.80(2) & 16.20(2) & 95.2(1) & 1834 \\
    ACRDIN10 & P2$_1$/c &  6.03(7) & 18.76(2) & 16.17(2) & 95.2(1) & 1821 \\
    ACRDIN11 & P2$_1$/n & 11.18(5) &  5.93(1) & 13.59(4) & 99.8(4) &  889 \\
    ACRDIN12 & P2$_1$/n & 11.28(1) & 12.38(1) &  6.68(1) & 92.1(1) &  933 \\
    \bottomrule
  \end{tabular}
\end{table}

Numerous theoretical studies have sought to determine the relative stability of acridine polymorphs.\cite{review_schur} However, classical force fields lack the flexibility required to accurately capture both the kinetic stability and the transitions between polymorphic forms.\cite{polymorphisme_md-qc} DFT, while more accurate, still exhibit lattice-energy errors ranging from approximately 4.18 to 20 kJ/mol depending on the functional employed, whereas experimental energy differences between polymorphs rarely exceed 10 kJ/mol.\cite{distribution_lattice_energy} 
The MLIP developed here is trained on DFT data to reproduce the PES, thereby enabling simulations at larger system sizes and finite temperatures to explore polymorphic stability.

\subsubsection{Dataset generation with LLMs}

By launching the first option of the AMLP codes by the following single prompt, "I am developing a machine‐learning interatomic potential for acridine polymorphs to predict energies, forces, and thermodynamic properties across different conditions." We were able to obtain four different reports that are in the SI. These reports allowed us to garner insight into what has been done experimentally and theoretically on this kind of crystal structure.It gives information on what functionals and basis sets could be used to study  crystalline forms for this system. Information regarding the importance of dispersion interactions is also noted.

Based on these considerations, we performed DFT calculations using the Perdew-Burke-Ernzerhof (PBE) functional~\cite{PBE} and a plane-wave basis set, as implemented in VASP, both of which were recommended by AMLP. In this particular case, employing a plane-wave basis set is preferable over Gaussian basis sets when investigating crystalline structures due to its better suitability using PBC. Note that, while plane-wave implementations of hybrid density functionals have traditionally been computationally demanding compared to their use in GGA functionals such as PBE, recent methodological advances have significantly improved the efficiency of evaluating hybrid functionals for periodic systems in a plane-wave basis\cite{mandal2022hybrid,mandal2019speeding,mandal2020efficient,mandal2018enhanced}, making the use of hybrid functionals more feasible for running simulations. The agents also suggested an electronic convergence threshold of \(10^{-6}\,\mathrm{eV}\), which we adopted. For the plane-wave cutoff, AMLP proposed 500~eV; here, we employed a more tighter value of 850~eV to ensure accurate total-energy evaluations. With respect to Brillouin-zone sampling, no explicit recommendation was provided by the agents, and we therefore selected a \(\Gamma\)-centered \(7 \times 7 \times 7\) Monkhorst--Pack mesh.\cite{monkhorst} For the treatment of long-range dispersion interactions, AMLP advised at least the use of D3(BJ) corrections\cite{d3bj}; however, we opted for Grimme’s D4 scheme,\cite{grimm-d4} which achieves accuracy comparable to the Many-Body Dispersion (MBD) method\cite{mbd-tatch1,mbd-tatch2,mbd-tatch3} while being substantially more efficient. The D4 correction ensures a reliable description of van der Waals interactions, which are essential for capturing the intermolecular forces governing crystal stability. All computational parameters are detailed in the SI. The crystal structures used in this study are presented in Table \ref{tab:acr_polymorphs}.

The DFT calculations of the eight polymorphs were subsequently processed using the AMLP code. The framework stores all relevant information in a structured \texttt{.json} format, as described in step 3 of Section \ref{sec:datapreparation}.

\begin{figure}[H]
    \centering
    \includegraphics[width=1.0\linewidth]{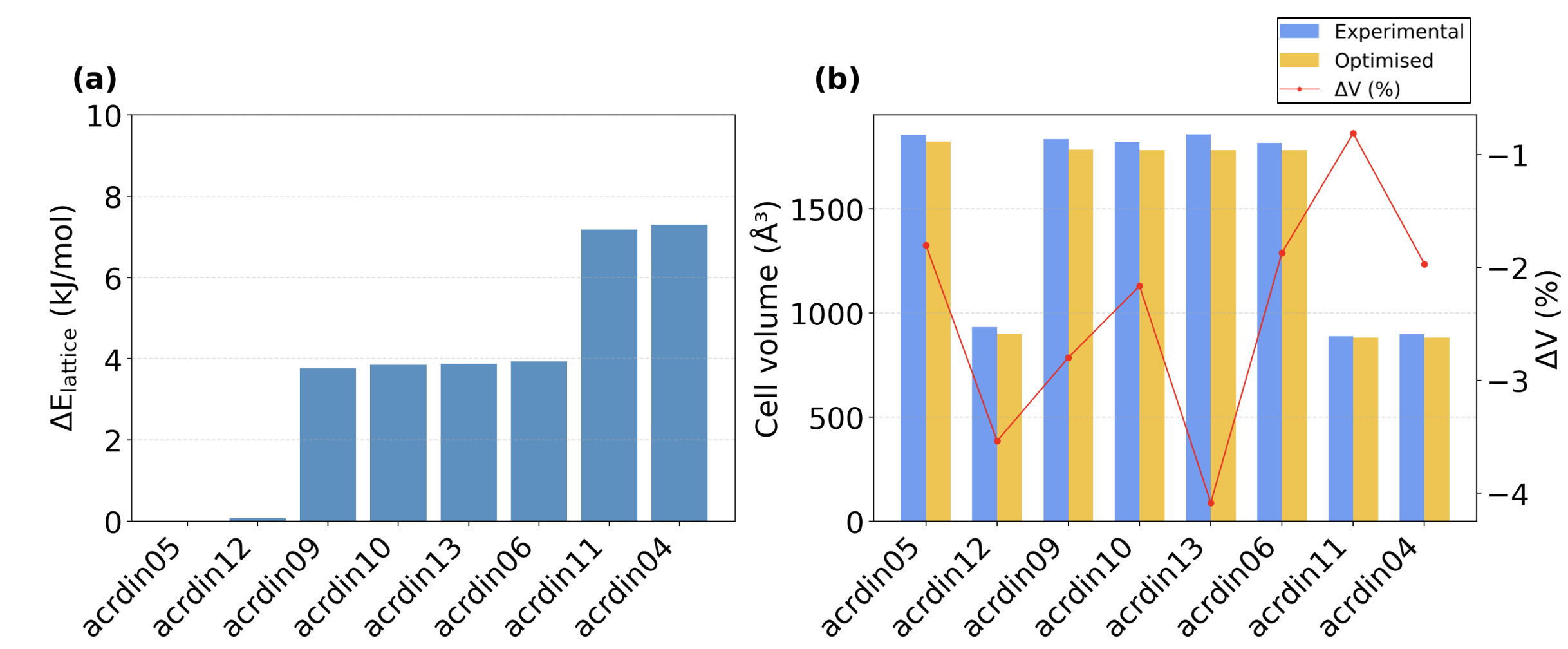}
    \caption{(a) $\Delta E$ denotes the difference in lattice energy between each polymorph and the most stable one (ACRDIN05).(b) Comparison between experimental (blue) and DFT optimized (yellow) unit-cell volumes of various acridine polymorphs. The red line corresponds to  $\Delta V$. }
    \label{fig:relaxedft-label}
\end{figure}

The reliability of the DFT calculations is evaluated by comparing the optimized unit-cell parameters and lattice energies against the corresponding experimental data from the CSD. The use of \texttt{.json} files further facilitates this process: by organizing data through easily readable keys, AMLP allows straightforward extraction and direct comparison of computed and experimental values. The DFT optimizations, performed at $T=0$~K, result in a contraction 
of the unit cell volume by approximately 3\% relative to the experimental structures 
determined near room temperature, which is expected due to the absence of thermal expansion.
The DFT optimized and experimental volumes as well as relative difference
\begin{equation}
\text{$\Delta V$} = \frac{V_{\mathrm{exp}} - V_{\mathrm{theo}}}{V_{\mathrm{exp}}}.
\end{equation}  
is shown together with the lattice energies

\begin{equation}
E_{\mathrm{lattice}} = \frac{E_{\mathrm{crystal}}}{Z} - E_{\mathrm{gas}},
\end{equation}  
in Figure~\ref{fig:relaxedft-label}.

By examining the lattice-energy differences relative to the most stable polymorph (ACRDIN05), we find that the overall range is approximately 7.3 kJ/mol. As noted previously, even though ACRDIN05 appears as the lowest-energy form in our DFT calculations, this does not conclusively indicate that it is more stable than the others, since this range of lattice energies falls within the intrinsic error of DFT methods. Notably, the energy difference between ACRDIN05 and ACRDIN12 is only $\sim$0.1~kJ/mol. 
A second group of four polymorphs (ACRDIN09, ACRDIN10, ACRDIN13, and ACRDIN06) has lattice energies higher by approximately 3.5~kJ/mol, only separated by 0.1 kJ/mol  from one another. The lattice energies of the last two polymorphs, ACRDIN11 and ACRDIN04 are even higher by another 3~kJ/mol and again differ by only 0.1 kJ/mol. These results, together with the observed \(3\%\) cell contraction, support the accuracy of our optimized structures and their ability to reproduce experimentally relevant stability trends. 

Following the cell optimizations, we carried out AIMD simulations with VASP across a temperature range of 300–800 K. This temperature range was specified interactively by using AMLP, to capture both room-temperature and high-temperature dynamical behavior, thereby ensuring that the potential remains transferable even under conditions approaching or exceeding the melting temperature. Specifically, the code identifies the relevant \texttt{.json} file from the preceding cell-optimization step, extracts the most stable geometry, and prepares the corresponding AIMD setup. Users can specify the temperature range and additional simulation parameters (e.g., timestep, ensemble), upon which AMLP automatically generates the corresponding input files. The analysis modules also provide default settings, ensuring that meaningful simulations can be carried out even without extensive user intervention.

\begin{figure}[H]
    \centering
    \includegraphics[width=1.0\textwidth]{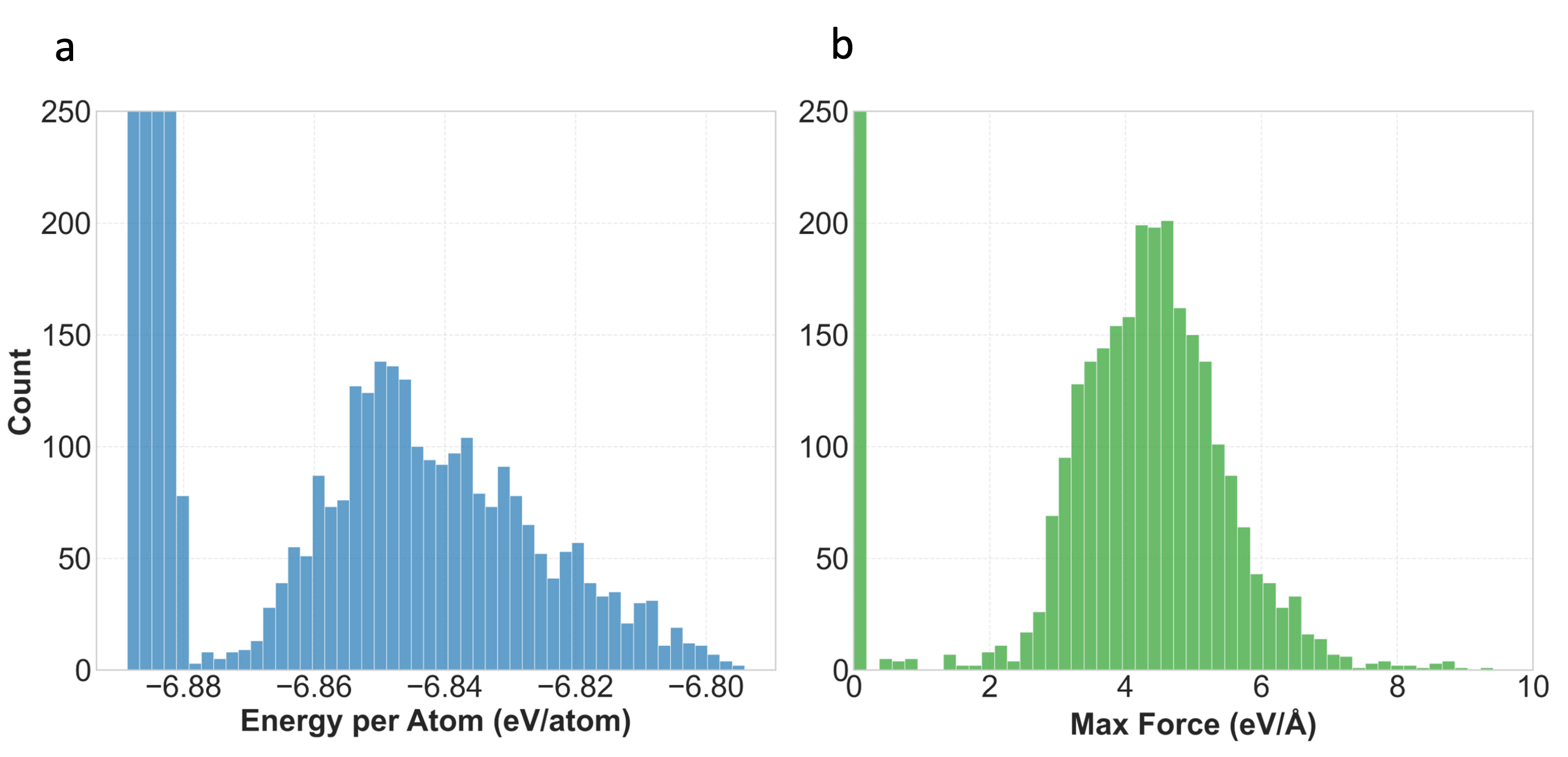}
    \caption{Distribution of the different structures coming from the DFT with (a) the energy per atom and (b) the maximum forces.}
    \label{fig:dataset}
\end{figure}

For the present work, we employed the canonical (NVT) ensemble with a timestep of 1.0~fs over 10,000 steps, using a Langevin thermostat to regulate the temperature. These parameters are user-defined and can be adjusted based on the specific requirements of each system. AMLP provides sensible default settings to serve as initial starting points for users. Additionally, users can consult the integrated LLM agents for system-specific parameter recommendations tailored to their particular molecular system. In total, AMLP automatically generated 58 simulation directories, each containing the required VASP input files (INCAR, POSCAR, and KPOINTS) for 8 polymorphs evaluated at 7 distinct temperatures. These directories were fully prepared for direct submission to the computing environment, thereby eliminating the need for manual intervention during the setup stage. The only action required from the user is to provide the initial POTCAR file. From each AIMD trajectory, several configurations per polymorph were extracted, resulting in a comprehensive dataset giving both equilibrium and non-equilibrium structures as depicted in Figure \ref{fig:dataset}. 

As with the previous steps, the outputs were automatically stored by the AMLP framework, resulting in a total of 8,208 structures. 

We observe a bimodal distribution, with equilibrium configurations characterized by forces in the range of 0-0.5 eV/\AA, and non-equilibrium configurations spanning approximately 2-10 eV/\AA. Before training, we employed the AMLP code to convert the generated \texttt{.json} files into the appropriate dataset format (HDF5). The dataset was then automatically partitioned into training and validation subsets, using an 85 \% to 15 \% split by default, with a batch size of 4. PBC were identified and applied automatically during this process. To improve data quality, a force cutoff of 8 eV/\AA\ was imposed to filter out outliers. After this preprocessing, a total of 8,108 structures remained available for model training.

\subsubsection{Computational Details}

Models were trained with the MACE implementation.\cite{MACE} At present, users manually specify all training hyperparameters. In future releases, we will integrate LLM agents to assist with parameter selection and configuration generation.
A cutoff radius of 6 \AA\ was employed to define the local environment of each atom, ensuring the inclusion of relevant neighbors while maintaining computational efficiency.

The log loss function was chosen for its balance between sensitivity to small deviations and robustness against larger errors. The MACE fine-tuning with the foundation model mace-mpa-0-medium.model has been used. We used a foundation model, as it has been trained on the whole periodic table and on materials, making our model more flexible. Moreover, training from this checkpoint would speed up the convergence of our fitting process.\cite{fine-tuning-juttafriend}. We trained the models for a total of 350 epochs in two stages. During the first stage (250 epochs), equal weights were assigned to the energy and force terms. In the second stage (100 epochs), the weight on the forces was increased to be ten times larger than that on the energies, in order to prioritize force fitting. For both stages of the training procedure, the same dataset was used. All details of the training parameters are provided in the SI.

\subsubsection{Validation and tests}

\begin{figure}[H]
    \centering
    \includegraphics[width=1.0\linewidth]{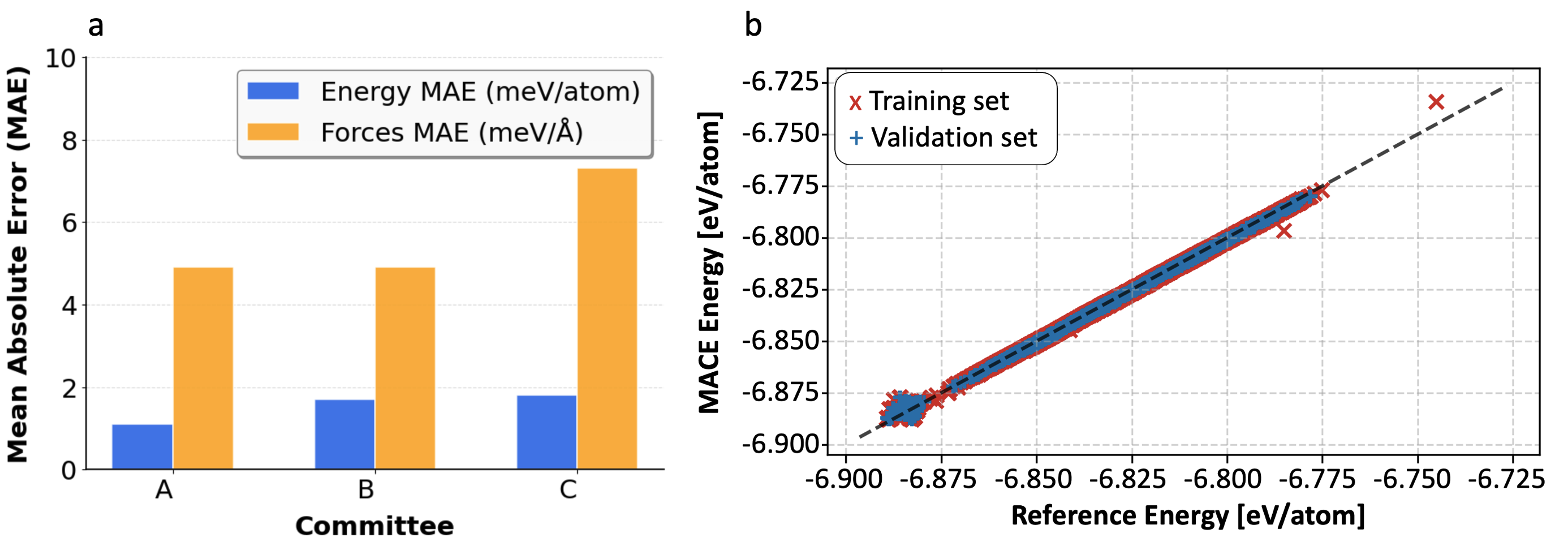}
    \caption{(a) Mean Absolute Error for the three different committees used. Comparing the MAE for energies in meV/atom (in blue) and for the forces in meV/Å (in yellow)}. (b) Correlation plots for MACE-A between the DFT reference and model energy per atom.
    \label{fig:mace-committee-MAE}
\end{figure}

To assess the robustness of our results, we trained three independent committees (MACE A/B/C) using distinct random seeds. Despite their independent initialization, all three committees converged to comparable levels of accuracy, yielding mean average errors (MAEs) of approximately $7~\text{meV/\AA}$ (equivalent to $0.67~\text{kJ/mol}$) for the forces and around $2~\text{meV/atom}$ ($0.19~\text{kJ/mol/atom}$) for the energies. These values, summarized in Figure~\ref{fig:mace-committee-MAE}, highlight both the consistency across the committees and the reliability of the training procedure in reproducing reference quantum-mechanical data. These quantities are automatically computed as part of the MACE training procedure.

The errors obtained for all committees remain well below the chemical accuracy threshold of 43.36 meV (1 kcal/mol), confirming the reliability of the models. For the relative forces, the three models achieve an average error of approximately \(\sim0.97\%\) on the validation sets, further demonstrating their accuracy. The correlation between the DFT reference data and the model predictions is high for energies, as illustrated for MACE-A in Figure~\ref{fig:mace-committee-MAE}.

Further validation is conducted by performing single-point energy calculations and geometry optimizations on the relaxed structures using each committee, and comparing the results with the corresponding DFT references. Subsequently, NVE simulations are carried out to assess energy conservation for each model. Finally, NVT simulations are performed to simulate all polymorphs under different thermodynamic conditions and check their integrities. This entire procedure was carried out using the AMLP-A module, which allows users to specify the desired sequence of simulations in a simple \texttt{config.yaml} file, enabling automated execution of the selected tasks in succession.

\subsubsection{Evaluation}

Recovering the exact lattice-energy order of polymorphs is inherently difficult, since the uncertainties of electronic-structure methods are of comparable or even greater magnitude. Consequently, the exact order cannot be determined with confidence even at the DFT level. We therefore focus on trends, identifying which machine-learned models most consistently preserve relative stabilities and yield usable energy landscapes, rather than enforcing exact agreement with a particular DFT ranking.

To this end, we analyzed all acridine polymorphs using the AMLP framework. Two complementary protocols were employed, both starting from DFT-optimized structures: (i) single-point (SP) energy evaluations to test how closely the fitted MLIP models reproduce DFT energies on the DFT geometries, and (ii) geometry optimization (OPT) using the MLIPs (force convergence threshold 10$^{-5}$ eV/\AA. Typical OPT wall-times were 2 min per polymorph on NVIDIA A100 GPUs.

\begin{figure}[H]
    \centering
    \includegraphics[width=0.8\linewidth]{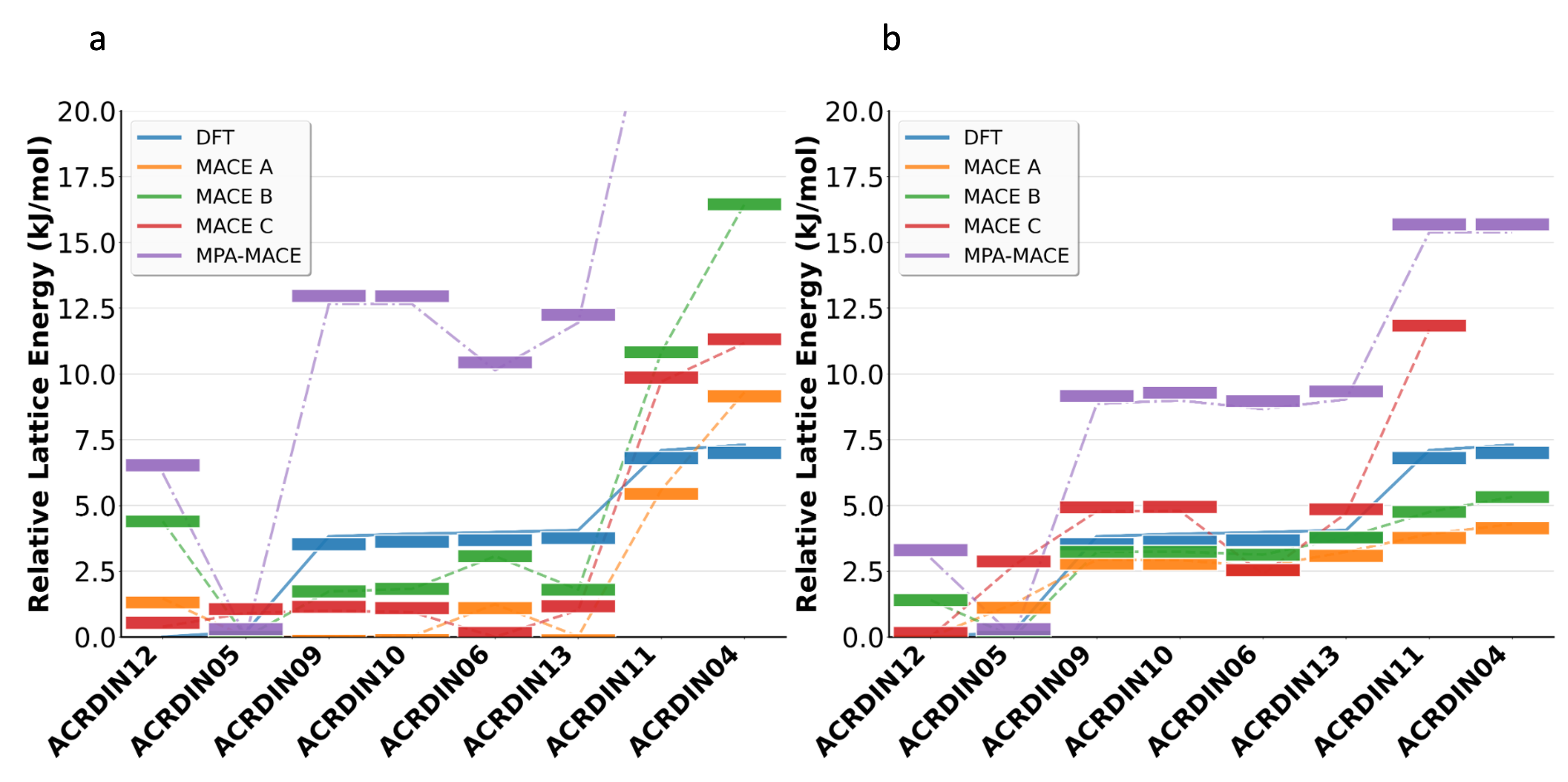}
    \caption{Relative lattice energies ($\Delta E_{\mathrm{lattice}}$) of acridine polymorphs predicted by the MPA–MACE foundation model and by fine-tuned committee models (MACE-A/B/C), compared against fully relaxed DFT references. (a): single-point evaluations on fixed geometries; (b): energies after geometry optimization.}
    \label{fig:mace_committee_stability}
\end{figure}

When comparing the relative lattice energies of the acridine polymorphs, all three committees reproduce the DFT energetic ordering reasonably well, with model-dependent quantitative offsets. By contrast, the untuned foundation model (purple curve in Fig. \ref{fig:mace_committee_stability}) deviates markedly; for example, it places ACRDIN04 at ~30 kJ/mol above the most stable form, which is larger than typical polymorphic separations. In single-point evaluations, the committees already recover the general trend, and subsequent geometry optimization further improves agreement with DFT for every model—including the MPA model—though the MPA model still exhibits larger relative-energy errors than the fine-tuned committees. Overall, these fine-tuned MLIPs better capture the shape of the polymorphic energy landscape (rank ordering and relative trends) than the absolute lattice energies.

Across the committees, deviations from DFT increase for higher-energy polymorphs—most notably ACRDIN11 and ACRDIN04—whereas low-lying structures are reproduced with smaller errors. MACE-C is consistently the least accurate, with pronounced discrepancies on the high-energy forms; for ACRDIN04 it did not converge to a physically meaningful minimum. Geometry optimization substantially reduces single-point discrepancies and improves rank ordering, indicating that AMLP fine-tuning of MACE foundation models robustly captures the polymorphic energy landscape.

To further assess structural fidelity, we examined the RMSD of atomic positions between DFT-relaxed structures and those obtained by geometry optimizations using the MACE models (Table \ref{tab:mace_rmsd}). RMSDs were calculated after optimal structural alignment using the Kabsch algorithm, which determines the best-fit rotation and translation to superimpose the structures before computing deviations.\cite{kabsh1,kabsh2} All models achieve mean RMSDs of $\sim0.048~\text{\AA}$, indicating that the predicted and reference structures are essentially identical. MACE-C is slightly more consistent, showing lower standard deviation and maximum deviation values.

\begin{table}[htbp]
\centering
\caption{RMSD of distances for fully DFT relaxed structures and geometry optimization for the different committees}
\label{tab:mace_rmsd}
\renewcommand{\arraystretch}{1.1}
\small
\begin{tabular}{l|cccc}
\toprule
MACE models & Mean (Å) & Std Dev (Å) & Min (Å) & Max (Å)\\
\midrule
MACE-A & 0.047 & 0.027 & 0.024 & 0.095 \\
MACE-B & 0.050 & 0.020 & 0.030 & 0.086 \\
MACE-C & 0.046 & 0.016 & 0.026 & 0.067 \\
\bottomrule
\end{tabular}
\end{table}

These results confirm that the three MACE committees reproduces DFT geometries with high fidelity, while capturing stability trends qualitatively even if absolute energy ranking differ.

At this stage, however, static evaluations (SP and OPT) only probe equilibrium structures and do not directly address the dynamical stability of the polymorphs under thermal conditions. A MLIP may reproduce plausible equilibrium structures while still producing unphysical trajectories when integrated over time. To establish whether the trained models generate stable and conservative dynamics consistent with their underlying potential-energy surfaces, it is therefore essential to test them under energy-conserving conditions.

Subsequently, to assess the dynamical robustness of the trained models, we performed 40 ps microcanonical (NVE) simulations with 1 ps of equilibration to evaluate energy conservation across different polymorphs. All models demonstrated excellent energy conservation in the range of $10^{-4}$ range for all polymorphs (Figure S1)
. Further details on the evaluation methodology and complete results are provided in the Supporting Information.

These results confirm that the MLIPs trained within our AMLP framework provide reliable energy conservation in microcanonical simulations. The committees establish a robust foundation for subsequent NVT simulations and analysis of radial distribution functions (RDFs), ensuring that observed structural and thermodynamic trends are not artifacts of poor energy conservation.

\begin{figure}[htbp]
  \centering
  \begin{overpic}[width=1.0\linewidth]{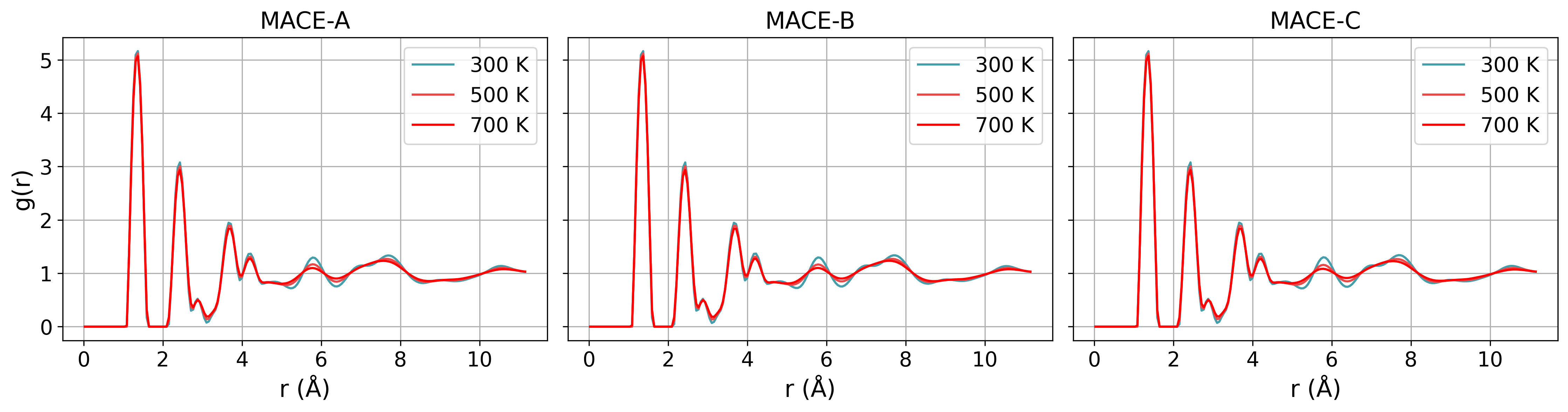}
    \put(2,26){\tiny\colorbox{white}{\textbf{(a)}}}
  \end{overpic}
  \vspace{0.5cm}
  \begin{overpic}[width=1.0\linewidth]{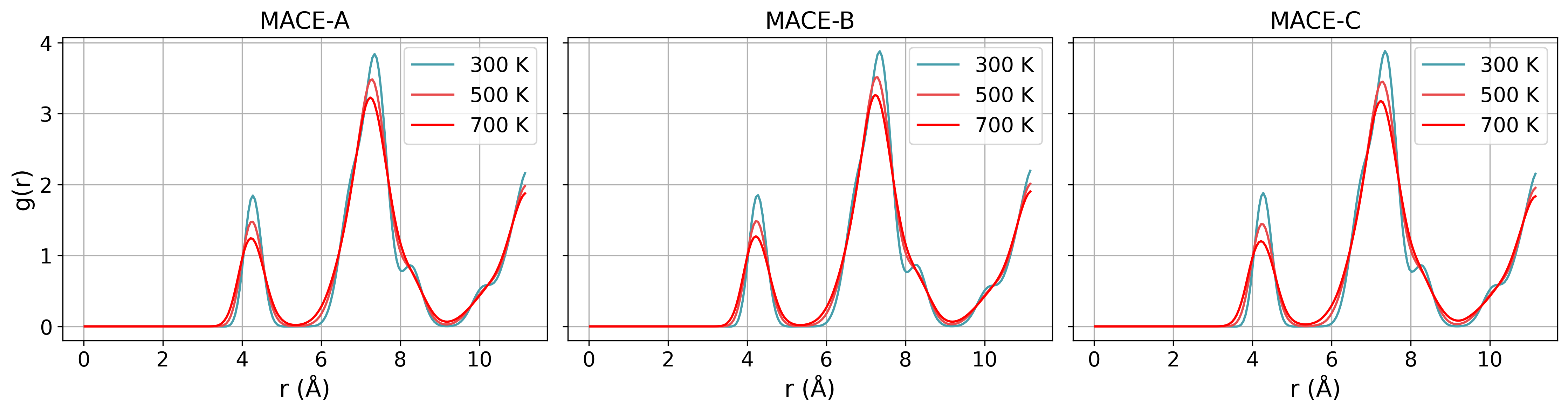}
    \put(2,26){\tiny\colorbox{white}{\textbf{(b)}}}
  \end{overpic}
  \caption{Radial distribution functions for ACRDIN12 at different temperatures and for different MLIP models. (a) C–N pairs. (b) N–N pairs.}
    \label{fig:acr12_rdf_combined}
\end{figure}

To investigate the structural stability of the different polymorphs, we performed NVT simulations using the different MLIP models. Within AMLP-A, users can select among various thermostats and integrators (see Section \ref{sec:amlpa}). At the end of each simulation, AMLP-a automatically computes RDFs for selected atomic pairs, thereby enabling a systematic analysis of the structural evolution with temperature. In particular, C–N pairs were analyzed as a probe of intramolecular stability, while N–N pairs served to characterize intermolecular packing interactions across a temperature range of $T = 300 - 700$~K. The AMLP analysis tool was used to automatically generate the RDF plots shown in Figure \ref{fig:acr12_rdf_combined}, enabling direct comparison between different models.

The RDF for C--N pairs in Figure \ref{fig:acr12_rdf_combined}(a) has distinct peaks at approximately 1.5 \AA\ and 2.3 \AA\ across all three models. The peak positions remain largely unchanged with increasing temperature, but their intensities decrease and the peaks broaden with increasing temperature, reflecting the expected increase in structural disorder and reduction of long-range correlations.

For N--N pairs (Figure~\ref{fig:acr12_rdf_combined}(b)), two sharp peaks at $\sim$4.1~\AA{} and $\sim$7.1~\AA{} dominate the RDFs. These features are consistently reproduced by all models and remain at fixed positions with temperature, showing the expected thermal broadening. At higher temperatures, minor secondary peaks (e.g., around 8.5~\AA{} and 10~\AA{}) disappear, indicating progressive destabilization of longer-range order.

Overall, both MACE-A and MACE-B reproduce consistent thermal broadening while preserving structural integrity, as expected for stable polymorphs (see SI for additional RDF graphs). In contrast, MACE-C exhibits severe instabilities: except for ACRDIN06 and ACRDIN12, all other polymorphs become unstable above room temperature, failing to maintain stable molecular packing under thermal excitation. Such behavior is absent in MACE-A and MACE-B, which display physically reasonable RDF trends across the full temperature range.

To assess the orientational order of molecular arrangements, we employed a $P_2$ order parameter analysis. This functionality is not yet integrated into AMLP, but will be included in future versions. This method analyzes all molecules within our simulation cells and calculates the orientational correlation between molecular planes according to:

\begin{equation}
P_2 = \left\langle \frac{3\cos^2\theta - 1}{2} \right\rangle
\end{equation}
where $\theta$ is the angle between molecular plane normal vectors, and the angular brackets $\langle \cdots \rangle$ indicate an ensemble average over all molecular pairs. The molecular plane normal vectors are determined using principal component analysis (PCA) of atomic coordinates within each molecule, where the normal corresponds to the eigenvector with the smallest eigenvalue.

The $P_2$ order parameter directly reflects molecular orientations: $P_2 = 1$ indicates perfect parallel alignment, $P_2 = 0$ corresponds to random orientations, and $P_2 = -0.5$ represents perfect perpendicular alignment. Values between $0.1-0.2$ typically indicate herringbone packing arrangements, while values above 0.5 suggest predominantly parallel packing. 

\begin{figure}[H]
    \centering
    \includegraphics[width=0.8\textwidth]{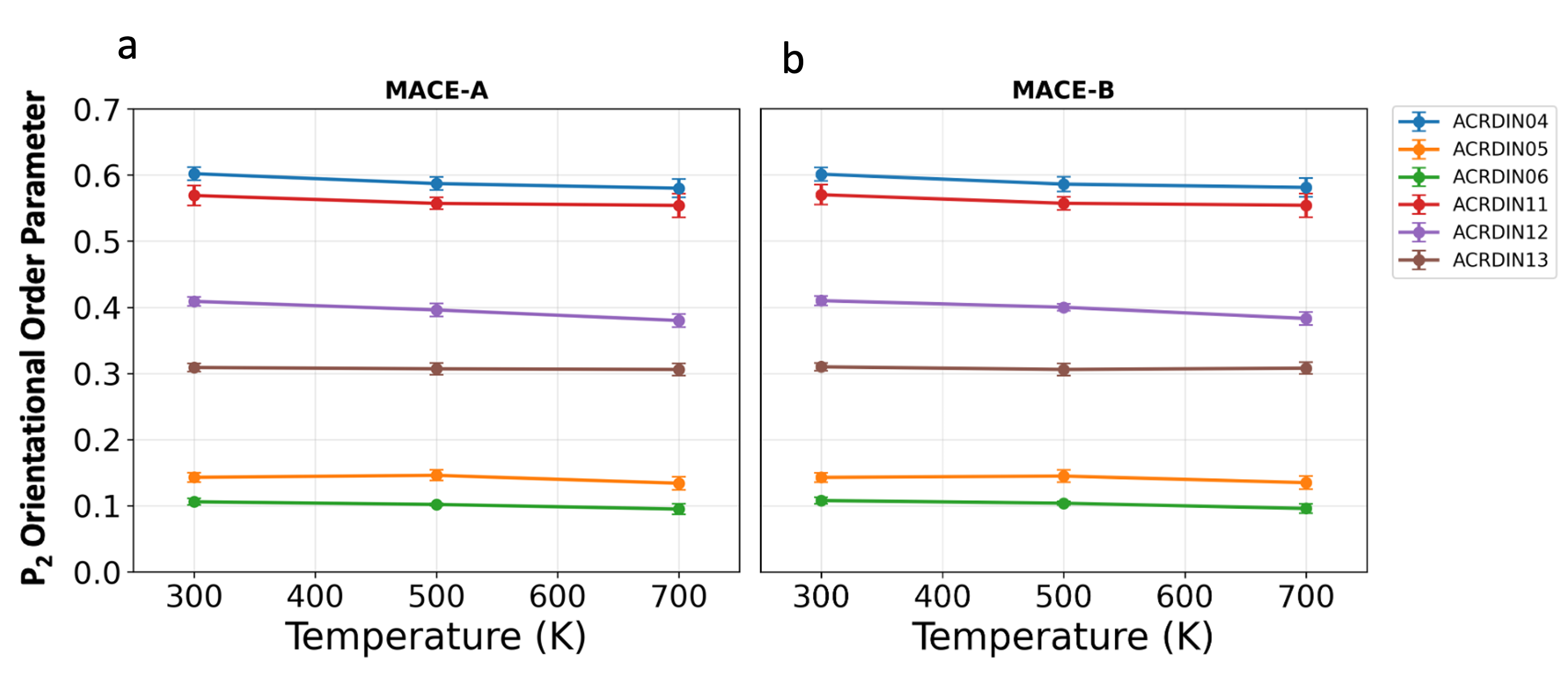}
    \caption{$P_2$ orientational order for acridine polymorphs at different temperatures using (a) MACE-A and (b) MACE-B potentials.}
    \label{fig:p2_heatmap}
\end{figure}

Figure~\ref{fig:p2_heatmap} demonstrates the thermal stability and molecular packing characteristics of six acridine polymorphs across the investigated temperature range. All polymorphs exhibit modest decreases in orientational order with increasing temperature, with \(P_{2} \) values typically decreasing by $0.01-0.02$  per 200~K increase, reflecting thermal fluctuations while maintaining crystalline integrity. ACRDIN04 and ACRDIN11 show the highest $P_{2}$ values ($0.58-0.60$), indicating predominantly parallel arrangements. ACRDIN12 and ACRDIN13 exhibit intermediate values ($0.38-0.41$), suggesting mixed packing motifs with both parallel and angled molecular arrangements. ACRDIN05 and ACRDIN06 display lower $P_{2}$ values ($0.14-0.15$), characteristic of herringbone packing where molecules adopt alternating orientations. Notably, all polymorphs maintain their characteristic packing motifs across the temperature range, demonstrating preserved crystalline character and thermal stability of the molecular arrangements. The excellent agreement between MACE-A and MACE-B potentials validates the reliability of both MLIPs for predicting orientational behavior and packing motifs in molecular crystals.

To assess the transferability of our MLIP to polymorphs outside the training set, we tested Form VIII, a recently characterized experimental structure that was not included in our dataset.\cite{review_schur} Geometry optimization and MD simulations were performed in both NVE and NVT ensembles. The optimized structures yielded relative lattice energies within an acceptable range: ~10.3 kJ/mol for MACE-A and MACE-B, and higher for MACE-C around ~17 kJ/mol (see Figure S12a).
The models also produced stable trajectories with proper energy conservation under the NVE ensemble (see Figure S12b) 
and stable dynamics under the NVT ensemble at 300 K for MACE-A and MACE-B (see Figure S12c). 
However, simulations at higher temperatures exhibited instabilities, indicating that the PES requires refinement for this polymorph under elevated-temperature conditions.
This case study illustrates both the strengths  and current limitations of the trained MLIPs. While the models shows reasonable transferability to unknown forms at low to moderate temperatures, configurations far from the training distribution—such as those accessed at high temperatures—require additional reference data. Crucially, AMLP facilitates this refinement process: structures from the unstable MD trajectories can be easily extracted, and the pipeline helps the users to  generate the necessary single-point calculations or AIMD simulations to augment the training set. This demonstrates the practical value of the automated framework for iterative MLIPs development and extension to new chemical environments.

\section{Conclusion}

In this work, we introduced AMLP, an automated machine learning pipeline designed to streamline the generation, training, and validation of MLIPs. The framework integrates LLM-driven agents for systematic dataset construction and incorporates fine-tuning procedures based on the MACE architecture.

We demonstrated the approach on acridine, a molecule with multiple known polymorphs that differ only by subtle structural variations. These simulations were computationally intensive, requiring significant CPU and GPU resources for cell optimizations, AIMD simulations, and model training. 

Our results show that the MLIPs automatically fine-tuned/trained with AMLP, achieves high accuracy, yielding close agreement with DFT references. Relative lattice energies is in agreement with DFT, rank-based stability trends are preserved. To further improve long-range accuracy, we plan to incorporate latent Ewald summation corrections into AMLP, which explicitly address long-range electrostatics.\cite{long-rangecorrection-ML}

Importantly, AMLP enables systematic validation of trained MLIPs against DFT geometries, which are reproduced with sub-\AA{} accuracy, confirming structural fidelity. The framework further automates dynamical assessments: under NVE conditions, trajectories remain stable and conservative with energy conservation better than 0.01\%. Subsequent NVT simulations and analyses demonstrate that AMLP provides a consistent framework for evaluating structural correlations across a wide range of temperatures. This enables a direct assessment of model quality, particularly in terms of their robustness in molecular dynamics simulations, and reveals distinct differences in performance between the MLIPs.

These findings emphasize the importance of dynamical validation, as equilibrium accuracy alone may be insufficient to guarantee model robustness at finite temperature. AMLP simplifies this entire process, lowering the barrier of entry for researchers who are new to machine-learned interatomic potentials.

The current AMLP framework represents a first complete version including the most important functionalities for an automated fine-tuning and fitting of MLIPs. Future developments will focus on expanding usability, for instance, by enabling LLM agents to directly generate plots or analyze \texttt{.json} datasets on demand, as recently demonstrated for agentized scientific workflows.\cite{DREAMSDensityFunctional2025} We also plan to extend support beyond MACE to other state-of-the-art models such as NequIP\cite{nequip}, TorchMD\cite{TorchMD}, and FeNNol\cite{fennol}, thereby broadening AMLP’s applicability and robustness in computational materials science and theoretical chemistry. In addition, the forthcoming release will introduce an automated AL orchestrator that translates user-specified objectives into cycling through inexpensive configuration generation with the use of different foundation models, selective high-level labeling, model retraining, and validation.

\section*{Acknowledgments}

\subsection*{Author Contributions} 
Adam Lahouari (ORCID: 0000-0001-5857-1066), Jutta Rogal (ORCID:0000-0002-6268-380X) and Mark E. Tuckerman (ORCID: 0000-0003-2194-9955) conceived and designed the
study. Adam Lahouari developed the code, performed the computational experiments,
analyzed the data, and drafted the manuscript. Jutta Rogal and Mark E. Tuckerman 
contributed to the writing and revision of the manuscript. All authors
gave final approval for the version to be published.

\subsection*{Funding}
This work was supported by the National Science Foundation grant DMR-2118890. 
JR acknowledges financial support from the Deutsche Forschungsgemeinschaft (DFG) through the Heisenberg Programme project 428315600. The Flatiron Institute is a division of the Simons Foundation. This work was supported in part through the NYU IT High Performance Computing resources, services, and staff expertise.

\subsection*{Conflicts of Interest}
The authors declare no competing interests.

\subsection*{Data Availability}
All data are available on the github of Adam Lahouari at \href{https://github.com/adamlaho/AMLP}{https://github.com/adamlaho/AMLP} 

\section*{Supporting Information}

Multi-agent system outputs including key findings and computational recommendations; VASP parameters for cell optimization; AIMD control parameters and electronic structure settings; MACE machine learning potential training parameters and atomic reference energies; energy conservation analysis demonstrating simulation stability; radial distribution functions characterizing local structure for all atomic pairs; out-of-sample polymorph validation results.

\bibliographystyle{achemso}
\bibliography{sample}
\clearpage

\textbf{\Large TOC Graphic}

\begin{figure}[H]
    \centering
    \includegraphics[width=0.75\linewidth]{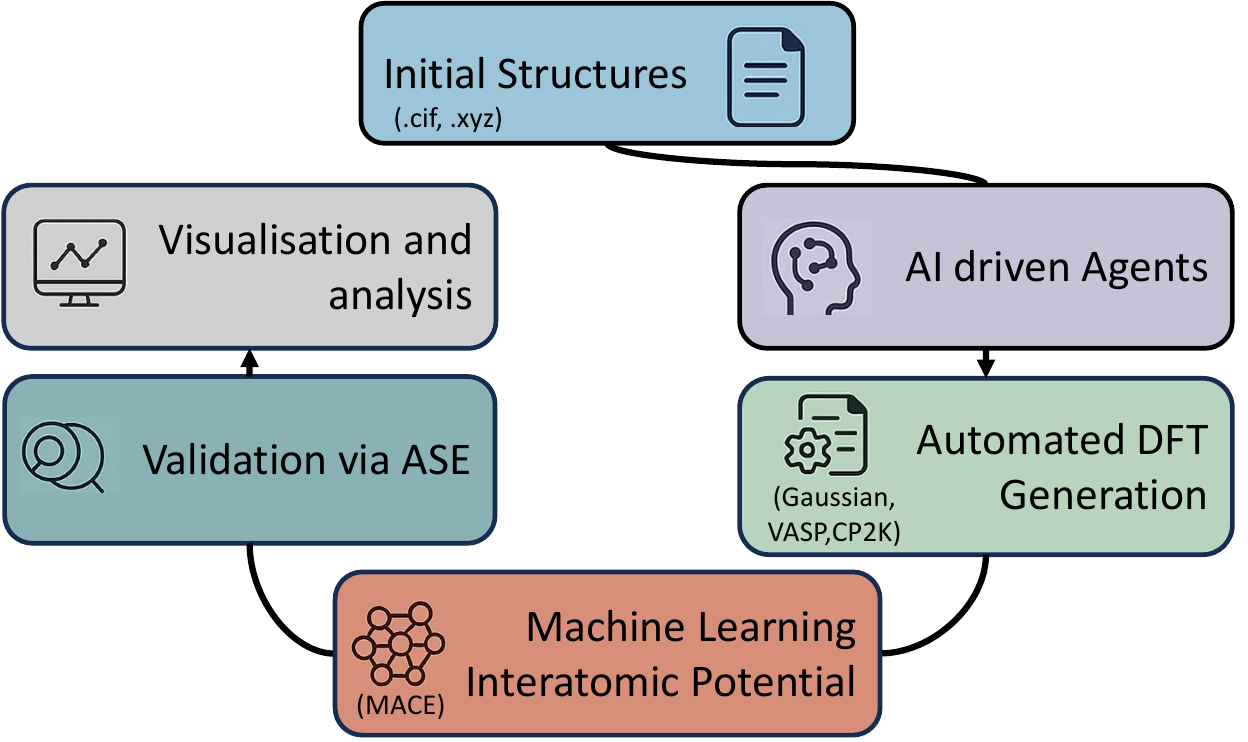}
    \label{fig:TOC}
\end{figure}

\clearpage

\appendix

\input{SI.tex}

\end{document}

%% file: SI.tex
\DeclareSIUnit\angstrom{\text{Å}}
\renewcommand{\thetable}{S\arabic{table}}
\setcounter{table}{0}
\DeclareSIUnit\angstrom{\text{Å}}

\renewcommand{\thepage}{S-\arabic{page}}
\setcounter{page}{1}

\begin{center}
    {\Large \textbf{Supporting Information:}}
    
    \vspace{0.5cm}
    
    {\Large \textbf{Automated Machine Learning Pipeline: Large Language Models-Assisted Automated Dataset Generation for Training Machine-Learned Interatomic Potentials.}}
    
    \underline{Adam Lahouari}$^{1,*}$, Jutta Rogal$^{1,2}$, and Mark E. Tuckerman$^{1,3,4,5,6}$
    
    \vspace{0.5cm}
    
    {\small
        $^{1}$NYU, Department of Chemistry, New York, NY 10003, USA
        
        $^{2}$Initiative for Computational Catalysis, Flatiron Institute, NY 10010, USA
        
        $^{3}$NYU, Department of Physics, New York, NY 10003, USA
        
        $^{4}$Courant Institute of Mathematical Sciences, NYU, NY 10012, USA
        
        $^{5}$NYU-ECNU Center for Computational Chemistry, Shanghai 200062, China
        
        $^{6}$Simons Center for Computational Physical Chemistry, NYU, NY 10003, USA
        
        $^{*}$Correspondence: al9500@nyu.edu
    }
\end{center}
\vspace{1cm}

\clearpage

\section*{Key Findings from AMLP Literature Analysis}
        Table~\ref{tab:key-findings} summarizes the major insights identified by the multi-agent literature analysis, with direct links to the retrieved publications and expert assessments.

        \begin{table}[H]
            \centering
            \caption{Key findings from AMLP multi-agent literature analysis for acridine polymorph research.}
            \label{tab:key-findings}
            \adjustbox{width=\textwidth,center}{
                \begin{tabular}{@{}llp{6cm}@{}}
                    \toprule
                    \textbf{Finding Category} & \textbf{Key Insight} & \textbf{Source \& Evidence} \\
                    \midrule
                    Acridine Literature & Limited direct studies on acridine polymorphs & Experimental \& theoretical reports: no specific acridine polymorph literature found \\
                    \midrule
                    Crystal Structure Prediction & AlphaCrystal-II deep learning approach & \texttt{arXiv:2404.04810v1} -- distance matrix-based crystal prediction \\
                    \midrule
                    Experimental Techniques & PXRD for crystal characterization & \texttt{arXiv:2401.03862v3} -- end-to-end crystal structure prediction from PXRD \\
                    \midrule
                    Novel Analysis Methods & Inverse EXAFS Analysis (IEA) & \texttt{arXiv:2409.09693v1} -- comparative study with Demeter software \\
                    \midrule
                    Materials Discovery & Autonomous ML-driven discovery & \texttt{arXiv:2508.02956v1} -- multi-agent physics-aware reasoning \\
                    \midrule
                    Similar Systems & Aromatic heterocycle precedents & Theoretical report: anthracene, phenanthrene, quinoline as model systems \\
                    \midrule
                    Computational Gaps & DFT + ML integration needed & All expert reports: lack of systematic acridine polymorph studies \\
                    \midrule
                    Methodology Limitations & Self-interaction error in DFT & Gaussian expert: \texttt{arXiv:2412.18350v1} -- exchange-correlation functionals \\
                    \midrule
                    Software Capabilities & VASP for solid-state calculations & VASP expert: \texttt{arXiv:2508.07035v1} -- VASPilot automation platform \\
                    \midrule
                    Force Field Development & Neural network potential comparison & VASP expert: \texttt{arXiv:2304.10820v1} -- Matlantis vs VASP comparison \\
                    \bottomrule
                \end{tabular}
            }
        \end{table}

    \subsection*{Research Opportunity Assessment}
        Based on the literature analysis, AMLP identified several research opportunities:

        \begin{table}[H]
            \centering
            \caption{Research gaps and opportunities identified by AMLP for acridine ML potential development.}
            \label{tab:research-gaps}
            \adjustbox{width=\textwidth,center}{
                \begin{tabular}{@{}llp{6cm}@{}}
                    \toprule
                    \textbf{Gap Type} & \textbf{Specific Opportunity} & \textbf{Recommended Approach} \\
                    \midrule
                    Experimental Data & Acridine polymorph characterization & Systematic PXRD and thermal analysis studies \\
                    \midrule
                    Computational Benchmark & DFT method validation for acridine & Compare PBE, PBE-D3(BJ), hybrid functionals \\
                    \midrule
                    ML Architecture & Neural network potential training & Combine VASP DFT data with modern ML frameworks \\
                    \midrule
                    Phase Transition & Thermodynamic modeling & Free energy calculations using enhanced sampling \\
                    \midrule
                    Software Integration & Automated workflow development & Extend VASPilot-type approaches to polymorph screening \\
                    \midrule
                    Validation Framework & Experimental-computational synergy & Cross-validate ML predictions with crystal growth experiments \\
                    \bottomrule
                \end{tabular}
            }
        \end{table}

        \renewcommand{\arraystretch}{1.2}
        \setlength{\tabcolsep}{6pt}
        \begin{table}[H]
          \centering
          \caption{Actual AMLP DFT-parameter recommendations for acridine crystalline systems.}
          \label{tab:dft-recs-actual}
          \begin{tabular}{@{}llp{9cm}@{}}
              \toprule
              \textbf{Software} & \textbf{Parameter} & \textbf{Recommended Value \& Provenance} \\
              \midrule
              \multirow{3}{*}{VASP} & ENCUT & 500 eV (molecular systems literature) \\
               & XC functional & PBE (standard for organic crystals) \\
               & Method focus & Periodic DFT with dispersion corrections \\
              \midrule
              \multirow{3}{*}{Gaussian} & Method & B3LYP (organic chemistry standard) \\
               & Basis set & 6-31G(d,p) (balanced accuracy/cost) \\
               & Application & Molecular property benchmarking \\
              \midrule
              \multirow{3}{*}{CP2K} & XC functional & PBE (consistent with VASP) \\
               & Basis set & DZVP-MOLOPT-SR-GTH \\
               & Method & Mixed Gaussian/plane-wave (GPW) \\
              \midrule
              \multirow{2}{*}{General} & Dispersion & D3(BJ) corrections recommended \\
               & Convergence & \(10^{-6}\) eV electronic, \(10^{-3}\) eV\,\AA\(^{-1}\) forces \\
              \bottomrule
            \end{tabular}%
        \end{table}

\clearpage

\section*{Computational Methods}

    \subsection*{Density Functional Theory (DFT) Calculations}
        All DFT calculations were performed using the Vienna Ab initio Simulation Package (VASP) with the Projector Augmented Wave (PAW) method.\cite{vasp1,vasp2,vasp3,vasp4} In this study on acridine, we decided to use VASP with the following parameters:

    \subsubsection*{Cell Optimization Parameters}
            \begin{table}[H]
                \centering
                \caption{VASP parameters for cell optimization}
                \label{tab:cell-opt-params}
                \adjustbox{width=0.8\textwidth,center}{
                    \begin{tabular}{lll}
                        \toprule
                        Parameter & Value & Description \\
                        \midrule
                        SYSTEM & acridine\_cell & System identifier \\
                        ALGO & Normal & Electronic minimization algorithm \\
                        EDIFF & \num{1e-6} \si{\electronvolt} & Electronic convergence criterion \\
                        ENCUT & \SI{850.0}{\electronvolt} & Plane-wave cutoff energy \\
                        GGA & PE & Exchange-correlation functional (PBE) \\
                        IBRION & 2 & Ionic relaxation algorithm (conjugate gradient) \\
                        ISIF & 3 & Stress tensor calculation (cell shape and volume optimization) \\
                        ISMEAR & -1 & Smearing method (Fermi smearing) \\
                        IVDW & 13 & van der Waals correction method \\
                        KPOINTS & 7×7×7 & K-point mesh (Gamma-centered, automatic) \\
                        LCHARG & .TRUE. & Write charge density \\
                        LREAL & Auto & Real-space projection \\
                        LVDW & .TRUE. & Enable van der Waals corrections \\
                        LWAVE & .TRUE. & Write wavefunctions \\
                        NELM & 100 & Maximum electronic self-consistency steps \\
                        NSW & 150 & Maximum ionic relaxation steps \\
                        PREC & High & Precision level \\
                        SIGMA & \SI{0.05}{\electronvolt} & Smearing parameter \\
                        VDW\_R0 & \SI{12.0}{\angstrom} & van der Waals cutoff radius \\
                        \bottomrule
                    \end{tabular}
                }
            \end{table}

        \subsubsection*{Ab Initio Molecular Dynamics (AIMD) Parameters}
            For the AIMD, the following parameters were used for different temperature:

            \begin{table}[H]
                \centering
                \caption{Molecular dynamics control parameters for AIMD simulation at 300 K}
                \label{tab:md-control}
                \adjustbox{width=0.8\textwidth,center}{
                    \begin{tabular}{lll}
                        \toprule
                        Parameter & Value & Description \\
                        \midrule
                        IBRION & 0 & Molecular dynamics mode \\
                        NSW & 10,000 & Number of MD steps \\
                        POTIM & \SI{1.0}{\femto\second} & Time step \\
                        TEBEG & \SI{300.0}{\kelvin} & Initial temperature \\
                        TEEND & \SI{300.0}{\kelvin} & Final temperature \\
                        ISYM & 0 & Symmetry disabled for MD \\
                        MDALGO & 3 & Langevin thermostat \\
                        ISIF & 2 & Stress tensor calculation (ionic positions only) \\
                        \bottomrule
                    \end{tabular}
                }
            \end{table}

            \paragraph{Langevin Thermostat Parameters:}
                \begin{itemize}
                    \item \textbf{LANGEVIN\_GAMMA}: 10.0 10.0 10.0 (friction coefficients for \ch{N}, \ch{C}, \ch{H} atoms)
                    \item \textbf{LANGEVIN\_GAMMA\_L}: 1.0 (lattice friction coefficient)
                    \item Species-specific friction coefficients: \ch{N}: $\gamma = \SI{10.0}{\tera\hertz}$, \ch{C}: $\gamma = \SI{10.0}{\tera\hertz}$, \ch{H}: $\gamma = \SI{10.0}{\tera\hertz}$
                \end{itemize}

            \begin{table}[H]
                \centering
                \caption{Electronic structure and algorithm parameters for AIMD}
                \label{tab:aimd-electronic}
                \adjustbox{width=0.8\textwidth,center}{
                    \begin{tabular}{lll}
                        \toprule
                        Parameter & Value & Description \\
                        \midrule
                        ENCUT & \SI{850.0}{\electronvolt} & Plane-wave cutoff energy \\
                        EDIFF & \num{1e-6} \si{\electronvolt} & Electronic convergence criterion \\
                        NELM & 100 & Maximum electronic steps \\
                        ISMEAR & -1 & Fermi smearing \\
                        SIGMA & \SI{0.05}{\electronvolt} & Smearing parameter \\
                        GGA & PE & PBE exchange-correlation functional \\
                        LVDW & .TRUE. & Enable van der Waals corrections \\
                        IVDW & 13 & van der Waals method \\
                        VDW\_R0 & \SI{8.0}{\angstrom} & van der Waals cutoff radius \\
                        PREC & High & Precision level \\
                        ALGO & Fast & Electronic minimization algorithm \\
                        LREAL & Auto & Real-space projection \\
                        LWAVE & .FALSE. & Do not write wavefunctions \\
                        LCHARG & .FALSE. & Do not write charge density \\
                        NBLOCK & 1 & Output frequency \\
                        KBLOCK & 10 & k-point output frequency \\
                        LCHIMAG & .FALSE. & Magnetic moments output \\
                        \bottomrule
                    \end{tabular}
                }
            \end{table}

    \subsection*{Machine Learning Model Training}

        \subsubsection*{Data Preparation}
            For the training dataset preparation:
            \begin{itemize}
                \item \textbf{Force cutoff}: \(\SI{10.0}{\electronvolt\per\angstrom}\) (configurations with forces exceeding this threshold were excluded from training)
                \item \textbf{Atomic species}: \ch{H} ($Z=1$), \ch{C} ($Z=6$), \ch{N} ($Z=7$)
                \item \textbf{Data split}: 85\% training, 15\% validation
            \end{itemize}

        \subsubsection*{MACE Model Configuration}
            The Machine learning Atomic Cluster Expansion (MACE) model was trained using a foundation model approach with the following configuration:

            \begin{table}[H]
                \centering
                \caption{MACE model architecture and training parameters}
                \label{tab:mace-config}
                \adjustbox{width=0.9\textwidth,center}{
                    \begin{tabular}{lll}
                        \toprule
                        Parameter & Value & Description \\
                        \midrule
                        \multicolumn{3}{l}{\textbf{Foundation Model Settings}} \\
                        foundation\_model & mace-mpa-0-medium.model & Base pre-trained model \\
                        multiheads\_finetuning & False & Transfer learning approach \\
                        \midrule
                        \multicolumn{3}{l}{\textbf{Model Architecture}} \\
                        num\_channels & 128 & Number of hidden channels \\
                        max\_L & 2 & Maximum angular momentum \\
                        num\_interactions & 2 & Number of interaction layers \\
                        correlation & 3 & Correlation order \\
                        max\_ell & 3 & Maximum spherical harmonic degree \\
                        r\_max & \SI{6.0}{\angstrom} & Cutoff radius \\
                        \midrule
                        \multicolumn{3}{l}{\textbf{Training Parameters}} \\
                        device & cuda & Training device \\
                        batch\_size & 12 & Training batch size \\
                        max\_num\_epochs & 350 & Maximum training epochs \\
                        default\_dtype & float64 & Numerical precision \\
                        \midrule
                        \multicolumn{3}{l}{\textbf{Optimization Settings}} \\
                        swa & True & Stochastic Weight Averaging \\
                        start\_swa & 250 & SWA start epoch \\
                        swa\_lr & 0.0001 & SWA learning rate \\
                        swa\_forces\_weight & 10 & SWA forces weight \\
                        swa\_energy\_weight & 1 & SWA energy weight \\
                        ema & True & Exponential Moving Average \\
                        ema\_decay & 0.99 & EMA decay rate \\
                        amsgrad & True & AMSGrad optimizer variant \\
                        \midrule
                        \multicolumn{3}{l}{\textbf{Loss Function and Scaling}} \\
                        loss & ef & Energy and forces loss \\
                        scaling & rms\_forces\_scaling & Force scaling method \\
                        error\_table & PerAtomMAE & Error metric \\
                        \bottomrule
                    \end{tabular}
                }
            \end{table}

        \subsubsection*{Atomic Reference Energies}

            \begin{table}[H]
                \centering
                \caption{Atomic reference energies (E0s) used in MACE training}
                \label{tab:atomic-energies}
                \begin{tabular}{ll}
                    \toprule
                    Element & Energy (\si{\electronvolt}) \\
                    \midrule
                    \ch{H} & -0.27737421 \\
                    \ch{C} & -0.77091629 \\
                    \ch{N} & -0.83778712 \\
                    \bottomrule
                \end{tabular}
            \end{table}

        \subsubsection*{Ensemble Training}
            To ensure robustness and estimate model uncertainty, three independent models were trained using different random seeds with identical hyperparameters:
            \begin{itemize}
                \item \textbf{Seeds}: 49, 67, 127
                \item \textbf{Training epochs}: 300 epochs each
                \item \textbf{Committee approach}: All models trained with identical hyperparameters to form an ensemble for improved prediction accuracy and uncertainty quantification
                \item \textbf{Total number of parameters}: 752,174 
            \end{itemize}

\clearpage

\renewcommand{\thefigure}{S\arabic{figure}}
\setcounter{figure}{0}

\section*{Energy Conservation}

\begin{figure}[H]
    \centering
    \includegraphics[width=0.80\linewidth]{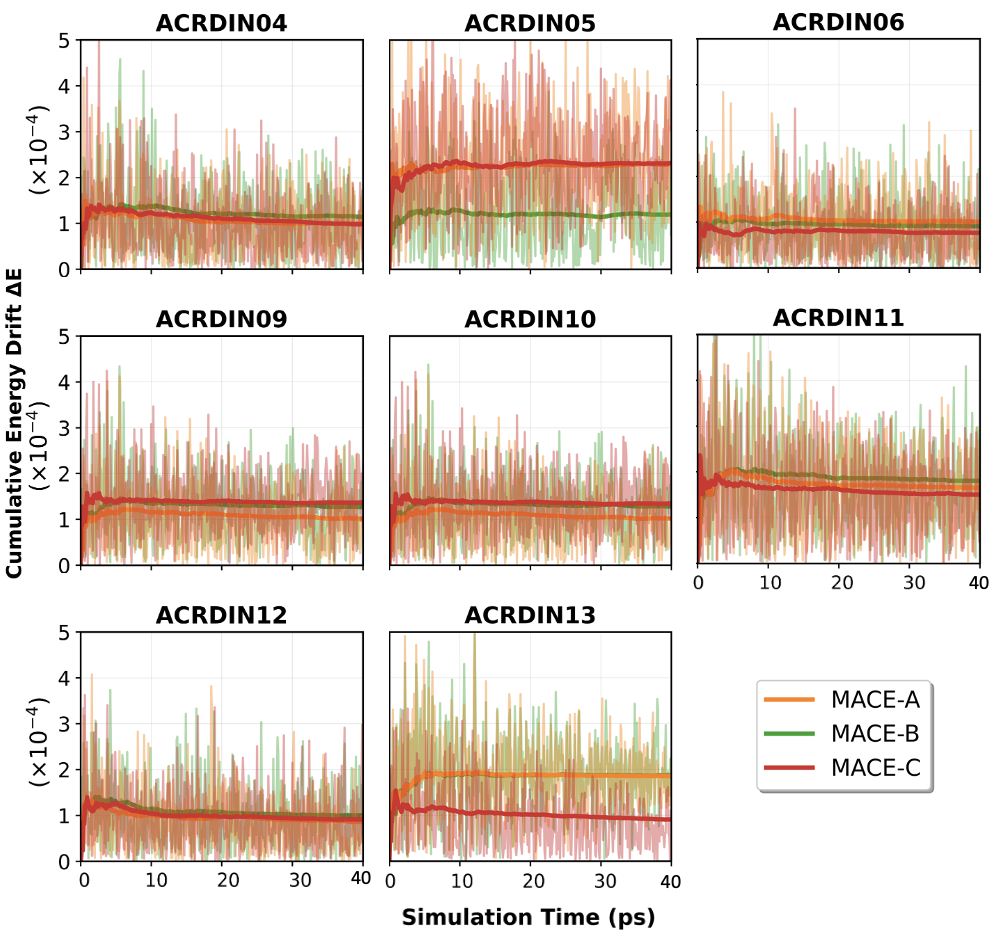}
    \caption{Cumulative energy conservation from NVE simulations of 39 ps following 1 ps of equilibration across all eight acridine polymorphs. Each panel corresponds to one polymorph, with results shown for three independently trained MLIPs: MACE-A (orange), MACE-B (green), and MACE-C (red). Solid lines represent the cumulative energy conservation, while shaded fluctuations reflect instantaneous deviations. All simulations show energy conservation on the order of $10^{-4}$.}
    \label{fig:energy-drift}
\end{figure}

The corresponding cumulative energies are summarized in Figure~\ref{fig:energy-drift}. For these simulations, all unit cells were replicated, reaching approximately 25~\AA{} in each direction and resulting in system sizes ranging from 1,104 to 2,944 atoms, depending on the polymorph. To achieve this, a cell-replication tool was implemented within the AMLP-A module, enabling users to easily apply scaling factors along specified directions.

Across all 24 simulations (3 committees $\times$ 8 polymorphs), 
the total energies are well conserved 
with an average energy drift consistently on the order of $10^{-4}$. This performance confirms that the trained potentials are robust for energy conservation.

Interestingly, minor variations across polymorphs are observed. For example, ACRDIN04, ACRDIN06, ACRDIN09, ACRDIN10, ACRDIN11, and ACRDIN12 consistently exhibited very good energy conservation across all committees. By contrast, MACE-B produced the lowest variation for ACRDIN05, while MACE-C was slightly better for ACRDIN13. These subtle differences suggest that structural features of individual polymorphs may influence energy conservation.

\section*{Supplementary Figures}

    \subsection*{Radial Distribution Functions: C--N pairs}

        \begin{figure}[H]
            \centering
            \includegraphics[width=1.0\textwidth]{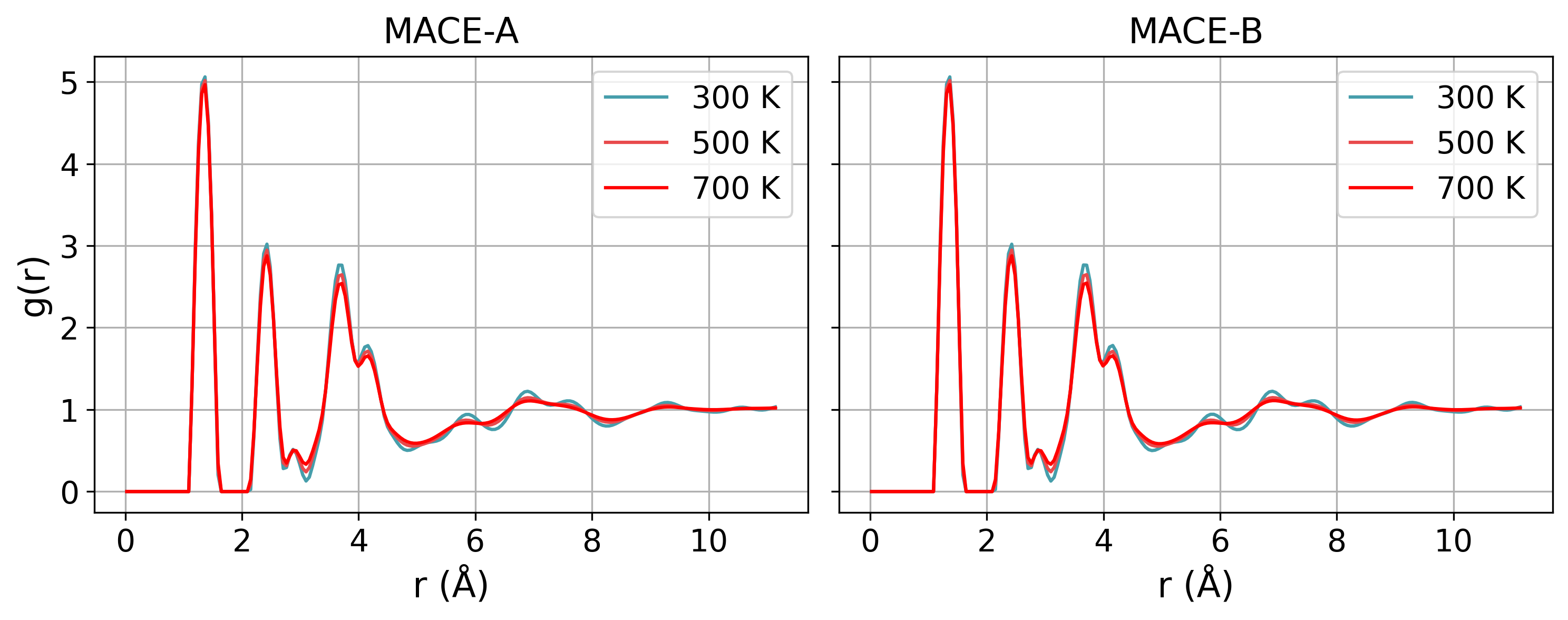}
            \caption{C--N pair radial distribution function $g_{\mathrm{C\!-\!N}}(r)$ for ACRDIN04 acridine polymorph.}
            \label{fig:rdf-cn-04}
        \end{figure}

        \begin{figure}[H]
            \centering
            \includegraphics[width=1.0\textwidth]{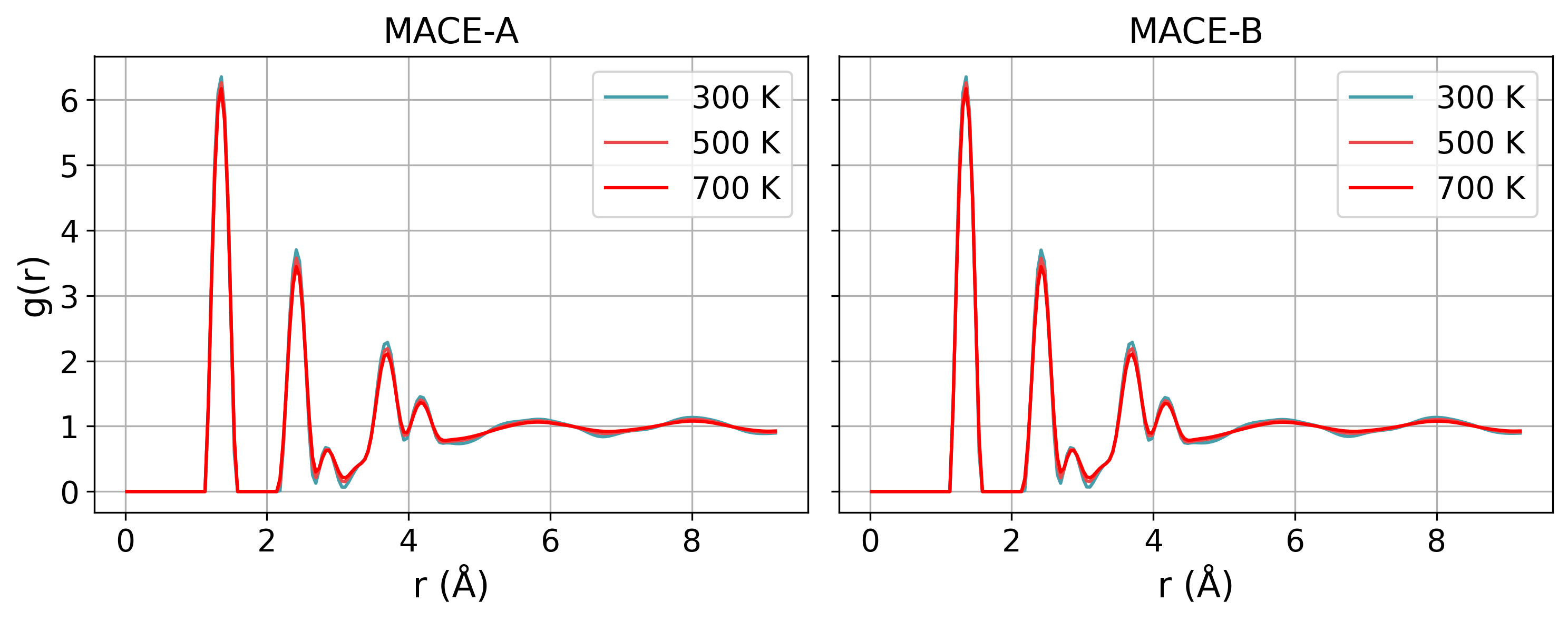}
            \caption{C--N pair radial distribution function $g_{\mathrm{C\!-\!N}}(r)$ for ACRDIN05 acridine polymorph.}
            \label{fig:rdf-cn-05}
        \end{figure}

        \begin{figure}[H]
            \centering
            \includegraphics[width=1.0\textwidth]{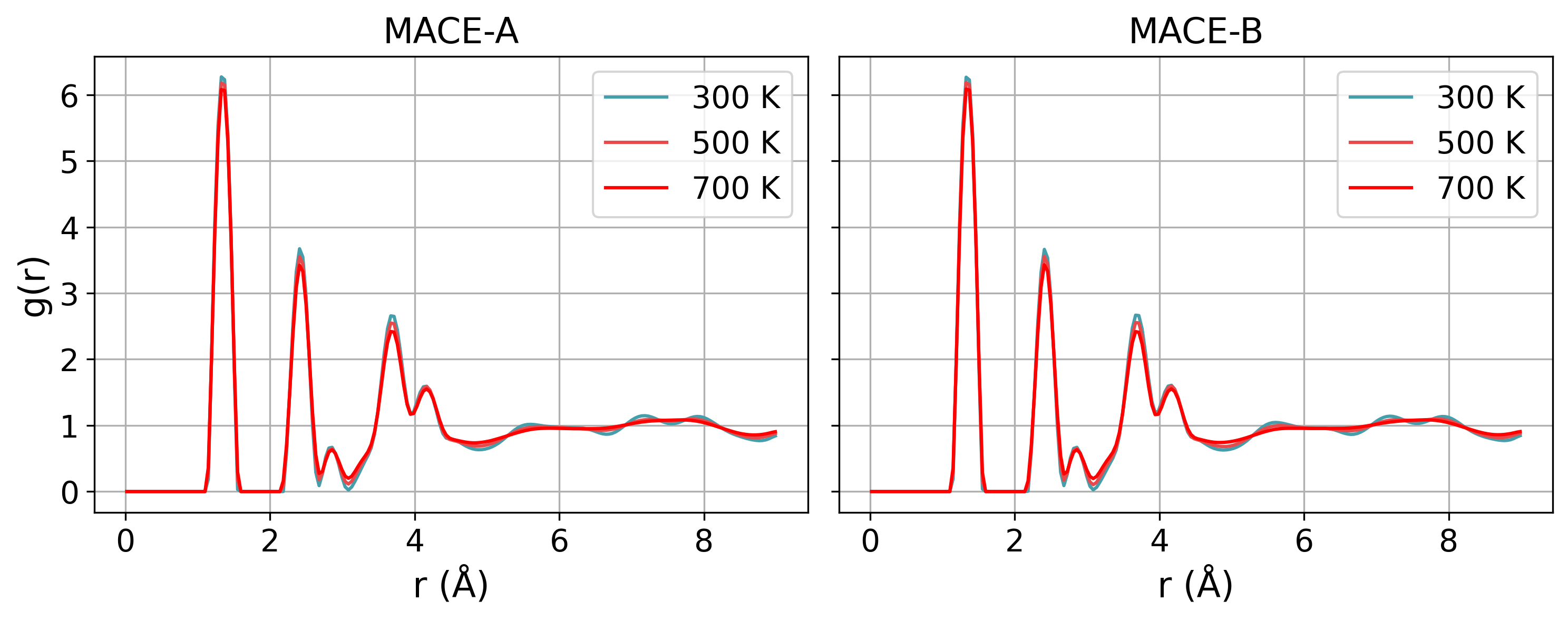}
            \caption{C--N pair radial distribution function $g_{\mathrm{C\!-\!N}}(r)$ for ACRDIN06 acridine polymorph.}
            \label{fig:rdf-cn-06}
        \end{figure}

        \begin{figure}[H]
            \centering
            \includegraphics[width=1.0\textwidth]{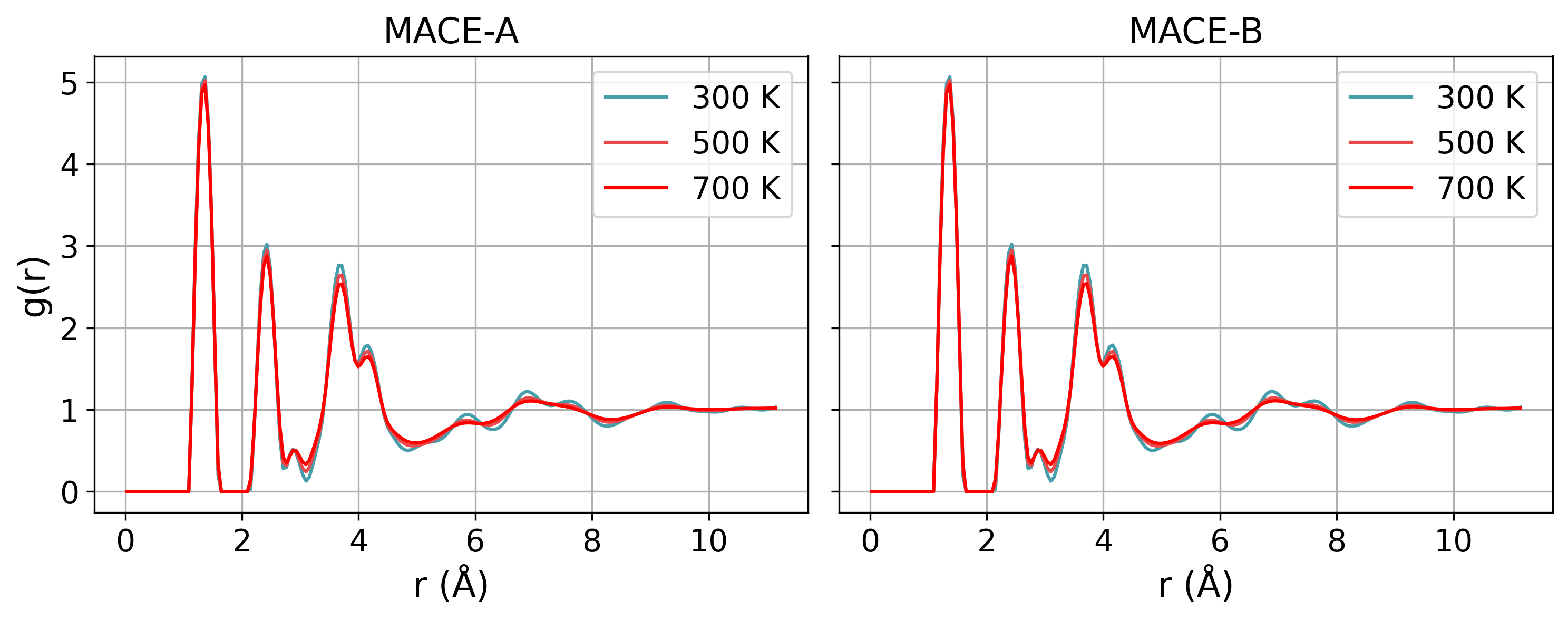}
            \caption{C--N pair radial distribution function $g_{\mathrm{C\!-\!N}}(r)$ for ACRDIN11 acridine polymorph.}
            \label{fig:rdf-cn-11}
        \end{figure}

        \begin{figure}[H]
            \centering
            \includegraphics[width=1.0\textwidth]{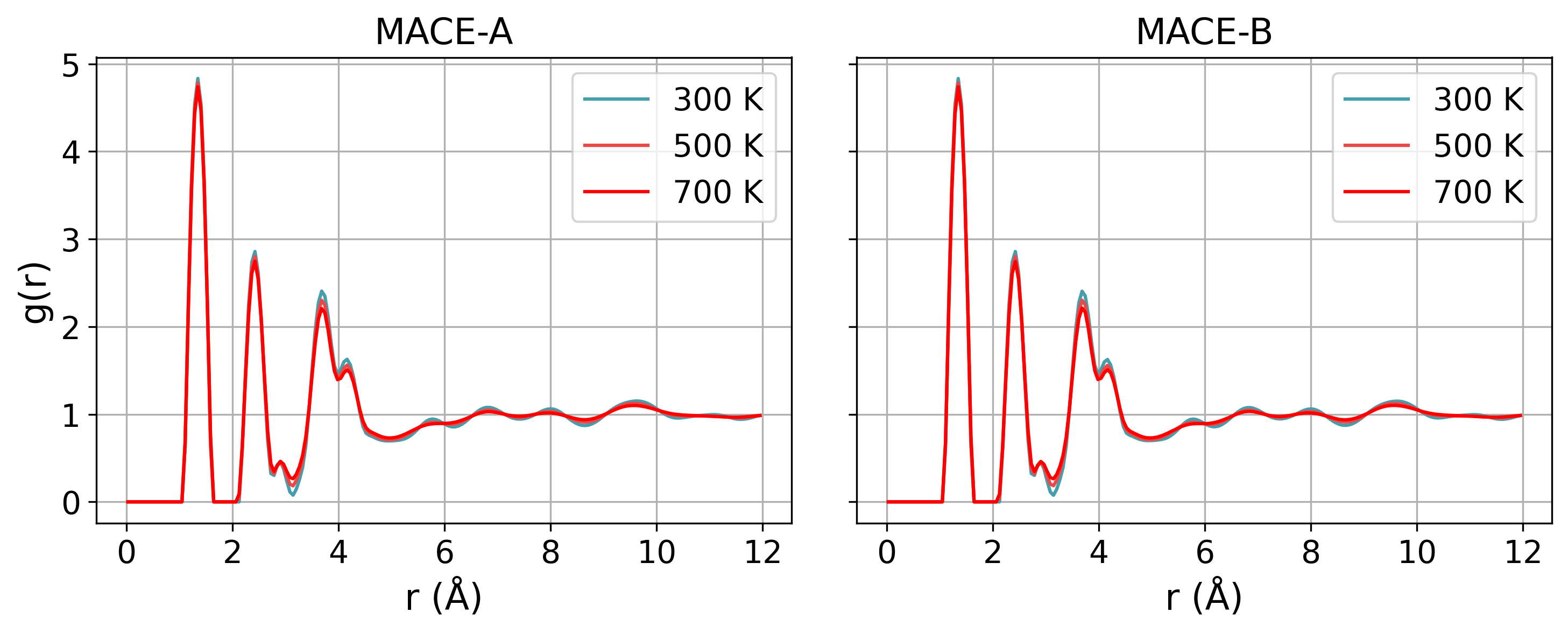}
            \caption{C--N pair radial distribution function $g_{\mathrm{C\!-\!N}}(r)$ for ACRDIN13 acridine polymorph.}
            \label{fig:rdf-cn-13}
        \end{figure}

    \subsection*{Radial Distribution Functions: N--N pairs}

        \begin{figure}[H]
            \centering
            \includegraphics[width=1.0\textwidth]{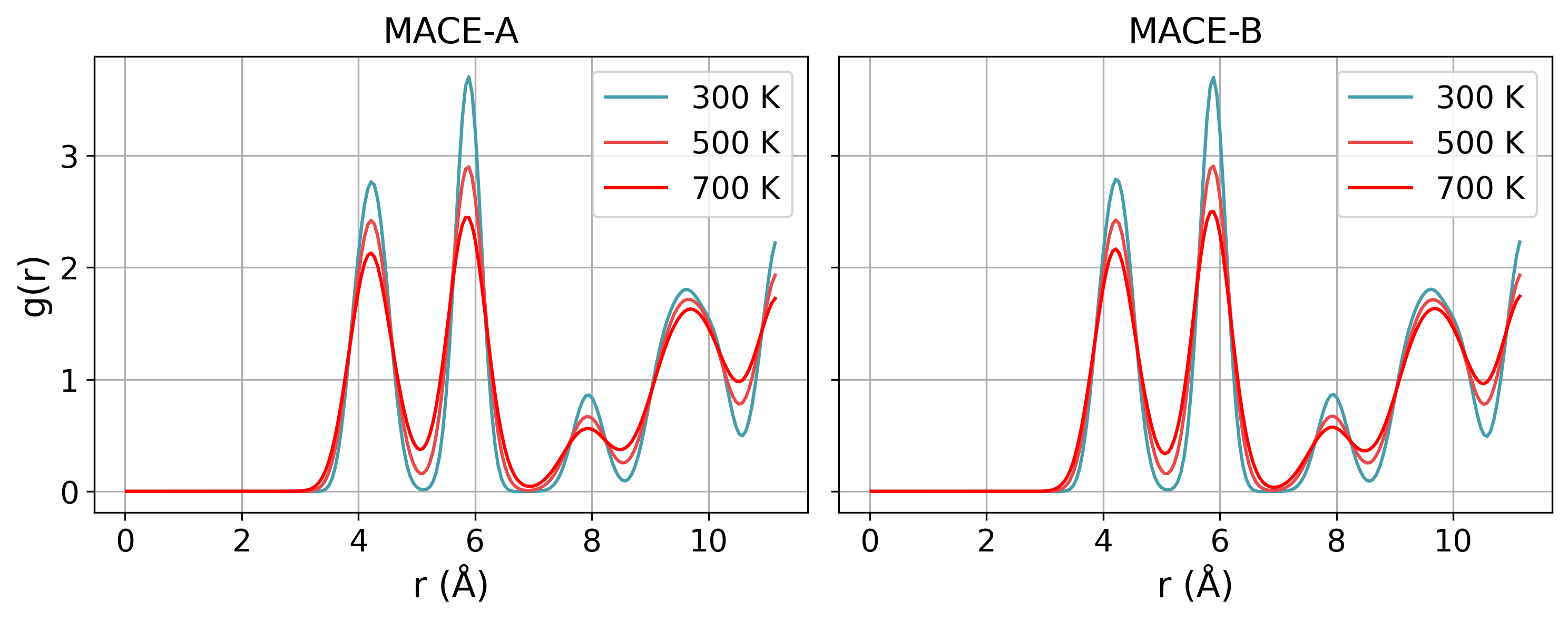}
            \caption{N--N pair radial distribution function $g_{\mathrm{N\!-\!N}}(r)$ for ACRDIN04 acridine polymorph.}
            \label{fig:rdf-nn-04}
        \end{figure}

        \begin{figure}[H]
            \centering
            \includegraphics[width=1.0\textwidth]{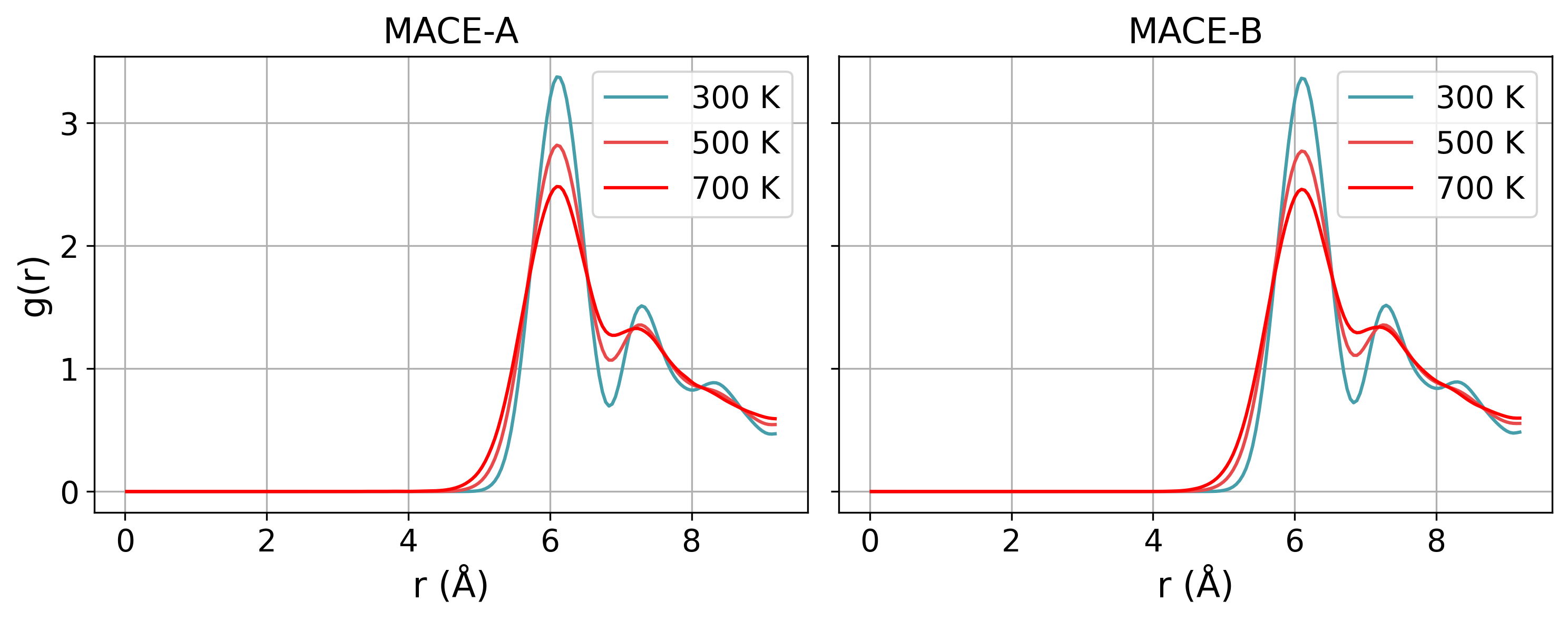}
            \caption{N--N pair radial distribution function $g_{\mathrm{N\!-\!N}}(r)$ for ACRDIN05 acridine polymorph.}
            \label{fig:rdf-nn-05}
        \end{figure}

        \begin{figure}[H]
            \centering
            \includegraphics[width=1.0\textwidth]{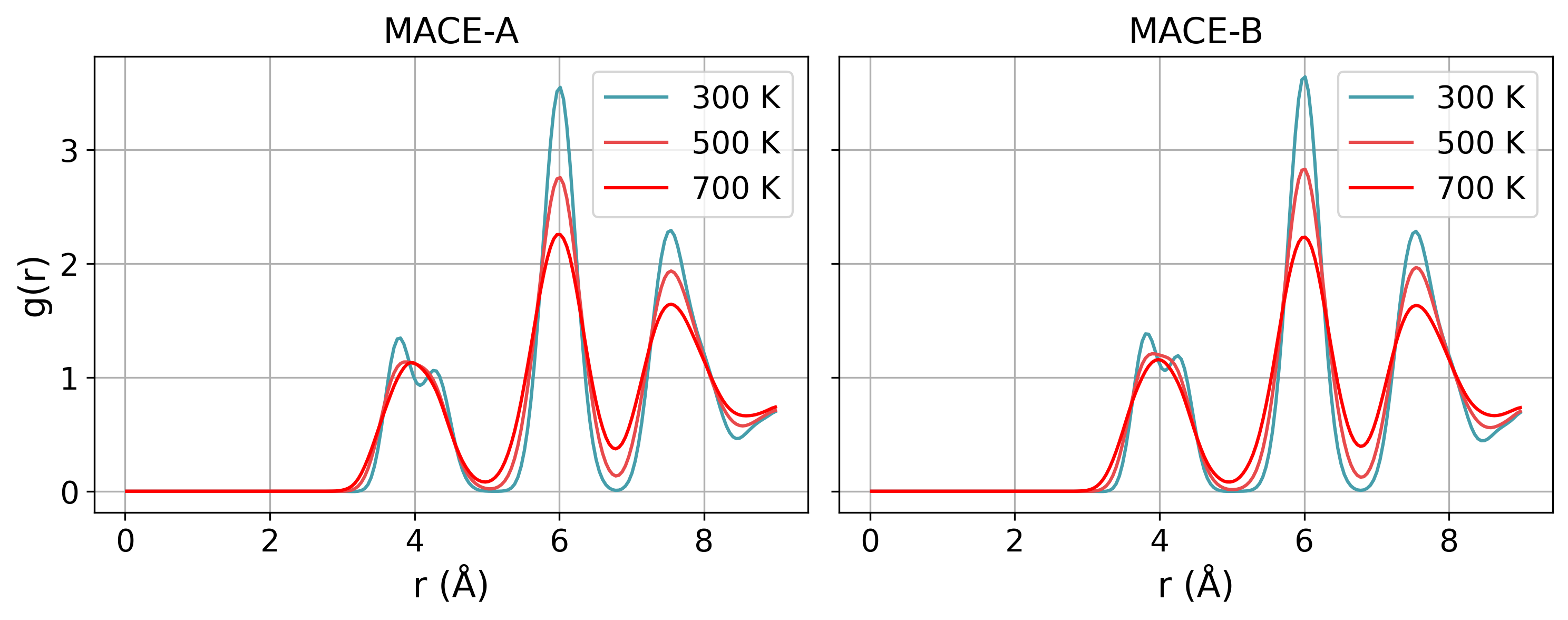}
            \caption{N--N pair radial distribution function $g_{\mathrm{N\!-\!N}}(r)$ for ACRDIN06 acridine polymorph.}
            \label{fig:rdf-nn-06}
        \end{figure}

        \begin{figure}[H]
            \centering
            \includegraphics[width=1.0\textwidth]{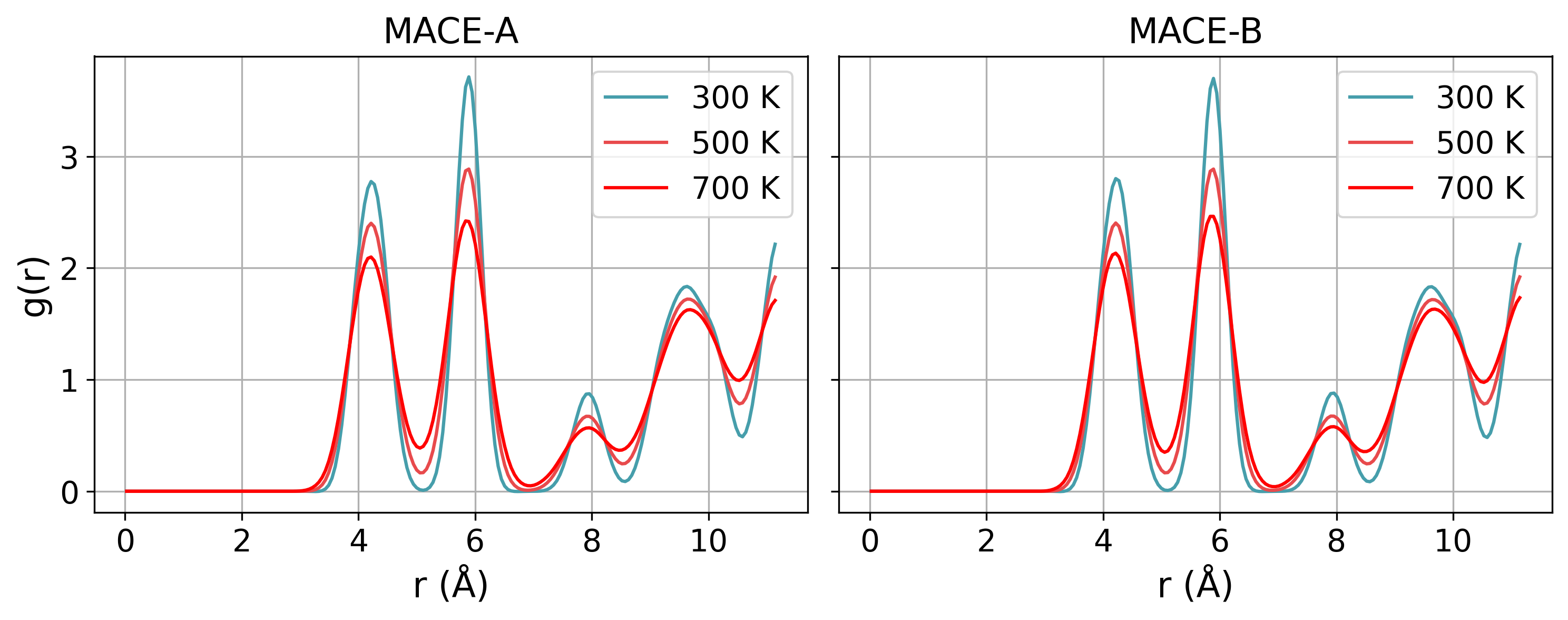}
            \caption{N--N pair radial distribution function $g_{\mathrm{N\!-\!N}}(r)$ for ACRDIN11 acridine polymorph.}
            \label{fig:rdf-nn-11}
        \end{figure}

        \begin{figure}[H]
            \centering
            \includegraphics[width=1.0\textwidth]{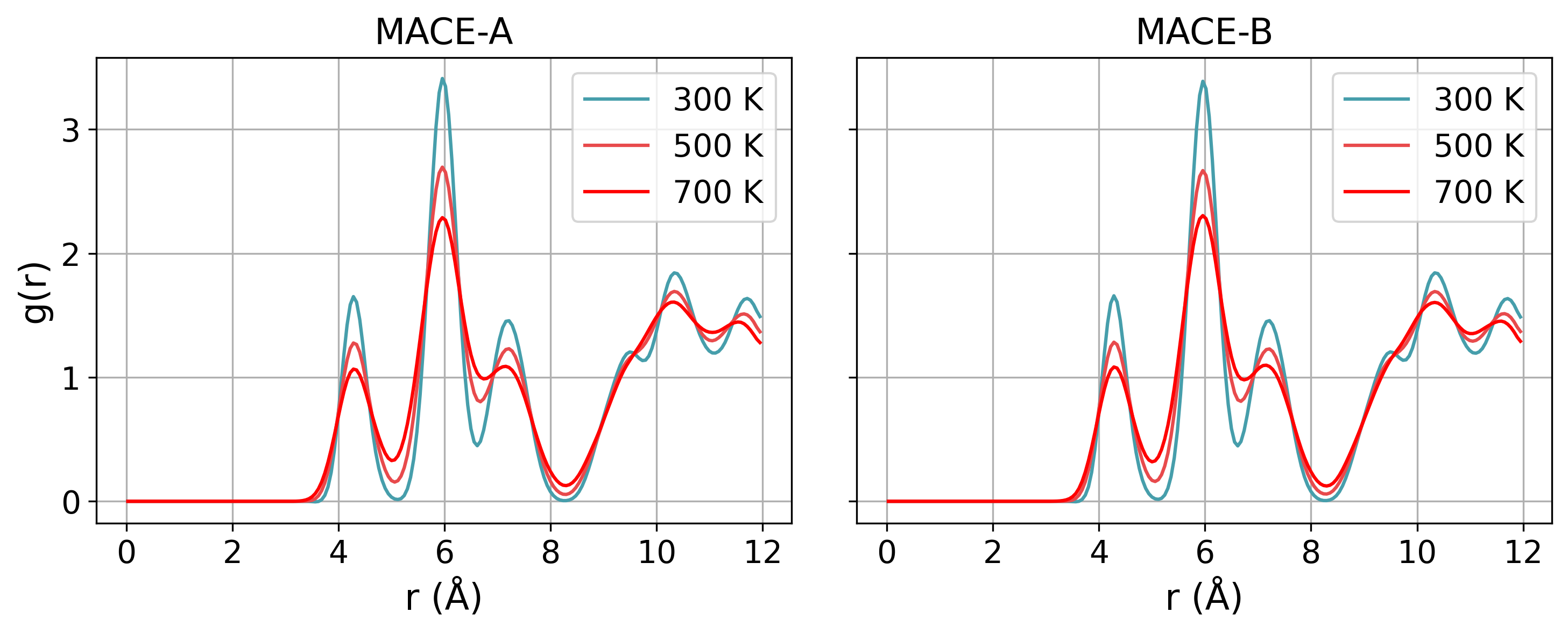}
            \caption{N--N pair radial distribution function $g_{\mathrm{N\!-\!N}}(r)$ for ACRDIN13 acridine polymorph.}
            \label{fig:rdf-nn-13}
        \end{figure}

\subsection*{Out-of-Sample Validation: ACRIDIN VIII}

\begin{figure}[H]
    \centering
    \includegraphics[width=1.0\linewidth]{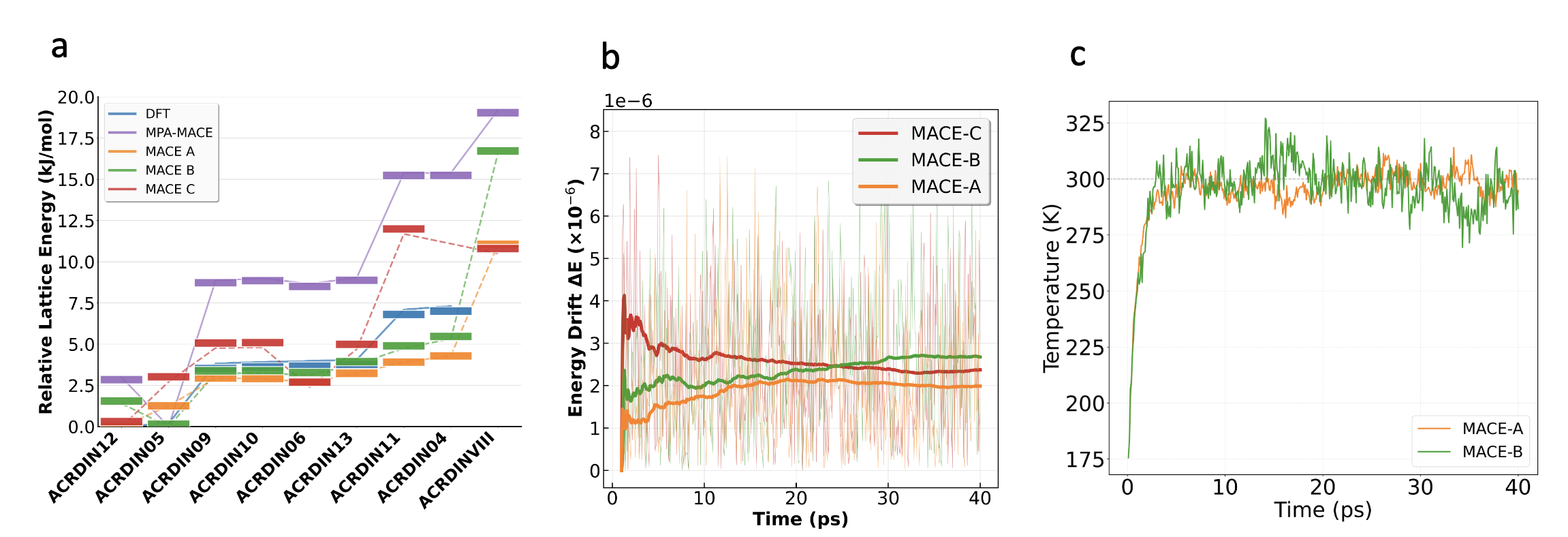}
    \caption{Performance evaluation of MACE committee models on ACRIDINVIII, a polymorph excluded from the training set. (a) Relative lattice energies of acridine polymorphs after geometry optimization, comparing DFT reference calculations with MPA-MACE and individual committee models (MACE-A, MACE-B, MACE-C). ACRIDINVIII (Form VIII) was not included in the training data. (b) Energy drift assessment during 40 ps NVE molecular dynamics simulations at 300 K using the three MACE committee models, demonstrating energy conservation with MACE-C showing the lowest drift (\(\Delta E ~ 2.0 \times 10^{-6}\) eV/atom). (c) Temperature stability during 40 ps NVT simulations at 300 K for MACE-A and MACE-B committees, showing equilibration around the target temperature with minimal fluctuations.}
    \label{fig:comparison-formVIII}
\end{figure}